\documentclass[lettersize,journal]{IEEEtran}
\usepackage{amsmath}
\usepackage{amsmath,amsfonts}
\usepackage{algorithmic}
\usepackage{algorithm}
\usepackage{array}
\usepackage{subeqnarray}
\usepackage{cases}
\usepackage{booktabs}
\usepackage{textcomp}
\usepackage{stfloats}
\usepackage{url}
\usepackage{verbatim}
\usepackage{stfloats}
\usepackage{graphicx}
\usepackage{multicol}
\usepackage{balance}
\usepackage{subcaption}
\usepackage{cite}
\usepackage{caption}
\usepackage{setspace}
\makeatletter

\newcommand{\Rmnum}[1]{\expandafter\@slowromancap\romannumeral #1@}
\makeatother
\usepackage{hyperref}
\hypersetup{hypertex=true,
            colorlinks=true,
            linkcolor=blue,
            citecolor=blue}

\begin{document}

\title{Monopulse Parameter Estimation based on MIMO-STCA Radar in the Presence of Multiple Mainlobe Jammings}

\author{\IEEEauthorblockN{
Huake Wang\IEEEauthorrefmark{1},~\IEEEmembership{Member,~IEEE}, Dongchang Zhang, Guisheng Liao,~\IEEEmembership{Senior Member,~IEEE}, and Yinghui Quan,~\IEEEmembership{Member,~IEEE}
\vspace{-0.25em}}

\thanks{This work was supported in part by the National Natural Science Foundation of China under Grants 62301410, in part by the Aviation Science Foundation under Grants 20230001081021, in part by the National Key Laboratory of Air-based Information Perception and Fusion under Grants ZH 2024-0053, and in part by the National Natural Science Foundation of China under Grants 62331019.}
\thanks{Huake Wang, Guisheng Liao and Yinghui Quan are with School of Electronic Engineering,
Xidian University, Xi'an, Shaanxi, China, 710071. Huake Wang, Dongchang Zhang, Guisheng Liao and Yinghui Quan are with Hangzhou Institute of Technology, Xidian University, Hangzhou, Zhejiang, China, 311231.}
}

% The paper headers

% Remember, if you use this you must call \IEEEpubidadjcol in the second
% column for its text to clear the IEEEpubid mark.

\maketitle

\begin{abstract}
The monopulse technique is characterized by its high accuracy in angle estimation and simplicity in engineering implementation. However, in the complex electromagnetic environment, the presence of the mainlobe jamming (MLJ) greatly degrades the accuracy of angle estimation. Conventional methods of jamming suppression often lead to significant deviations in monopulse ratio while suppressing MLJ. Additionally, the monopulse technique based on traditional radar cannot jointly estimate the target's range. In this paper, the four-channel adaptive beamforming (ABF) algorithm is proposed, which adds a delta-delta channel based on conventional sum-difference-difference three-channel to suppress a single MLJ. Moreover, considering the suppression of multiple MLJs and sidelobe jammings (SLJs), the row-column ABF algorithm is proposed. This algorithm utilizes more spatial degrees of freedom (DOFs) to suppress multiple jammings by the row-column adaptive beamforming at the subarray level. The key ideal of both algorithms is to suppress MLJ with null along one spatial direction while keeping the sum and difference beampatterns undistorted along another spatial direction. Therefore, the monopulse ratio remains undistorted while suppressing the MLJ, ensuring the accuracy of monopulse parameter estimation. Furthermore, by utilizing the additional degrees of freedom (DOFs) in the range domain provided by the multiple-input multiple-output space-time coding array (MIMO-STCA) radar, joint angle-range estimation can be achieved through the monopulse technique. Simulation results highlight the effectiveness of the proposed methods in suppressing multiple MLJs and enhancing the accuracy of monopulse parameter estimation, as verified by the low root mean square error (RMSE) in the parameter estimation results.

\end{abstract}


\section{Introduction}
\IEEEPARstart{T}{he} Monopulse is a well-established technique for radar angle estimation\cite{ref1},\cite{ref2},\cite{ref3}, which is widely employed in modern radar systems\cite{ref4}. The fundamental principle underlying the monopulse technique involves estimating the direction of the target by leveraging the ratio of the difference beam to the sum beam\cite{ref5}. However, in the context of modern battlefields, with the continuous upgrade of electronic defense systems, jamming techniques such as cross-eye jamming\cite{ref6}, angular glint jamming\cite{ref7}, and extremely high-power spot jamming\cite{ref8} have significantly degraded the performance of the monopulse technique, especially when dealing with jamming signals that are in close proximity to the center of the mainlobe.

In order to suppress the mainlobe jamming (MLJ), several methods based on adaptive beamforming were proposed in\cite{ref9},\cite{ref10},\cite{ref11}. However, due to the strong spatial correlation between the target and the MLJ, these methods often suppress the target signal alongside the MLJ, resulting in adverse effects such as mainlobe distortion, elevated sidelobe levels, and directional shifts. These issues lead to significant deviations in the monopulse ratio and a consequent reduction in parameter estimation accuracy\cite{ref12}. In\cite{ref13},\cite{ref14}, methods based on transmit polarization optimization have been proposed for suppressing MLJ. However, the effectiveness of these approaches may be significantly impacted by uncertainties in the electromagnetic environment. In\cite{ref15}, a mainlobe canceller employing adaptive digital beamforming was proposed for mainlobe jamming cancellation. However, this approach relies on precise alignment of the main beam with the target, which is challenging to accomplish in complex electromagnetic environments. A novel method for mainlobe jamming suppression leveraging the spatial polarization characteristics of antennas was proposed in\cite{ref16}. However, this approach relies on accurate estimation of the polarization state of the MLJ, otherwise the effectiveness of MLJ suppression will be significantly compromised. In \cite{ref17},\cite{ref18}, methods for MLJ suppression based on blind signal separation have been proposed, which can effectively separate the MLJ signal. However, these methods are associated with high computational complexity. For ground-based radar, a large auxiliary array was applied to suppress MLJ in\cite{ref19}. However, the method requires a large place to configure the antenna, which is the limitation in practical applications. A MLJ suppression approach based on covariance matrix reconstruction was introduced in\cite{ref20}. However, this method relies on prior knowledge of the MLJ's precise angle, which is challenging to obtain when limited training data is available. In\cite{ref21}, an adaptive monopulse approach with joint linear constraints on both the azimuth and elevation was proposed. This method flexibly adjusts the monopulse ratios for MLJ suppression at the subarray under constrained conditions, which is hard to determine.

As a widely known radar framework, the multiple-input multiple-output (MIMO) radar can transmit diverse mutually orthogonal waveforms simultaneously and employ multiple antennas to receive reflected signals, facilitating flexible spatial coverage\cite{ref22},\cite{ref23},\cite{ref24}. Compared to a standard phased array, waveform diversity offers several advantages, including higher resolution\cite{ref25}, better parameter identifiability\cite{ref26} and higher sensitivity to detecting slowly moving targets\cite{ref27}. As a novel transmit diversity technique, space-time coding array (STCA) has aroused great research interest nowadays\cite{ref28},\cite{ref29},\cite{ref30},\cite{ref31}. The fundamental concepts and characteristics of STCA were first proposed in\cite{ref32}, along with introducing fundamental tools and evaluation criteria for assessing STCA radar performance. As an innovative radar system built upon collocated MIMO radar, STCA introduces the time shift among transmit elements. This characteristic enables the beampattern of the STCA radar to be not only angle-dependent but also range-dependent. Regarding the application of STCA radar, \cite{ref34} proposed a beamspace multiple signal classification (MUSIC) algorithm to address the problem of direction-of-arrival (DOA) estimation in the case of small training samples. In\cite{ref35}, a simplified method based on the space-time coding principle was proposed, which simplifies the beampattern synthesis problem as a single waveform spectral shaping issue, enabling the generation of controllable transmit beampatterns with a lower computational complexity.

Notice that monopulse technique based on traditional radar cannot jointly estimate the target range. Combining existing MIMO technique\cite{ref36},\cite{ref37},\cite{ref38},\cite{ref39},\cite{ref40},\cite{ref41},\cite{ref42} with STCA technique, the MIMO-STCA radar can extract DOFs in the range domain through signal processing at the receive end. In this paper, the extra DOFs provided by the MIMO-STCA radar in the range domain are exploited to achieve joint angle-range estimation. Notice the problem that the presence of MLJ degrades monopulse performance, and conventional suppression methods often cause adverse effects. Consequently, utilizing the characteristic that the beampattern of a two-dimensional rectangular array is the product of independent row and column beampattern, the four-channel ABF algorithm is proposed. This algorithm suppresses MLJ with null along one direction while keeping the sum-and-difference beampatterns undistorted along another direction. First, set the time shift equal to 0, the MIMO-STCA radar degrades to a conventional MIMO radar, and the planar array becomes an azimuth-elevation two-dimensional planar array. Thus, MLJ suppression is applied along azimuth direction while preserving sum-and-difference beampatterns in elevation direction, ensuring unbiased elevation estimation. 
Similarly, we can achieve unbiased azimuth angle estimation. Second, set the time shift to a non-zero value and apply elevation angle dependence compensation to the vertical steering vector, transforming the MIMO-STCA radar into an azimuth-range two-dimensional planar array. Accordingly, we can perform unbiased estimation of target range. Moreover, the row-column ABF algorithm is proposed to suppress multiple MLJs and sidelobe jammings (SLJs) simultaneously. Compared to the four-channel ABF algorithm, the row-column ABF algorithm reverses the order of ABF and monopulse beamforming. Specifically, each row or column is utilized as a subarray for ABF, thus more DOFs are exploited to suppress multiple jammings. Then, monopulse beamforming is performed along column or row to jointly estimate angle and range.

The innovations proposed in this paper are summarized as follows:

1)A new joint angle-range estimation method for MIMO-STCA radar is proposed. Utilizing the extra DOF in range domain introduced by STCA radar, sum-and-difference beams in range domain are formed after ABF. Then, the ratio of the difference beam to the sum beam is used for range estimation.

2)A new method for MLJ suppression in monopulse parameters estimation is proposed. The key idea is canceling MLJ with null along one direction while keeping the sum-and-difference beampatterns undistorted along another direction. This approach can suppress MLJ while maintaining undistorted monopulse ratio for unbiased parameters estimation.

3)A new method for suppressing multiple MLJs and SLJs at subarray level is proposed. First, analog beamforming is performed at element level and subarray outputs are converted to digital signals. Then, adaptive row and column digital beams are formed at subarray level, utilizing the multiple DOFs to adaptively suppress multiple jammings.

\textit{Notations}: The remainder of this paper is organized as follows. Section \ref{section:2} derives the signal model for MIMO-STCA radar. To enhance the accuracy of monopulse parameters estimation in the presence of MLJ, the four-channel ABF algorithm is introduced in Section \ref{section:3}. Then, in Section \ref{section:4}, row-column ABF algorithm is proposed to suppress multiple jammings. The simulation results are given in Section \ref{section:5}, and conclusions are finally drawn in Section \ref{section:6}.

\section{MIMO-STCA Radar Signal Model}
\label{section:2}
\begin{figure}
    \centering
    \includegraphics[width=\columnwidth]{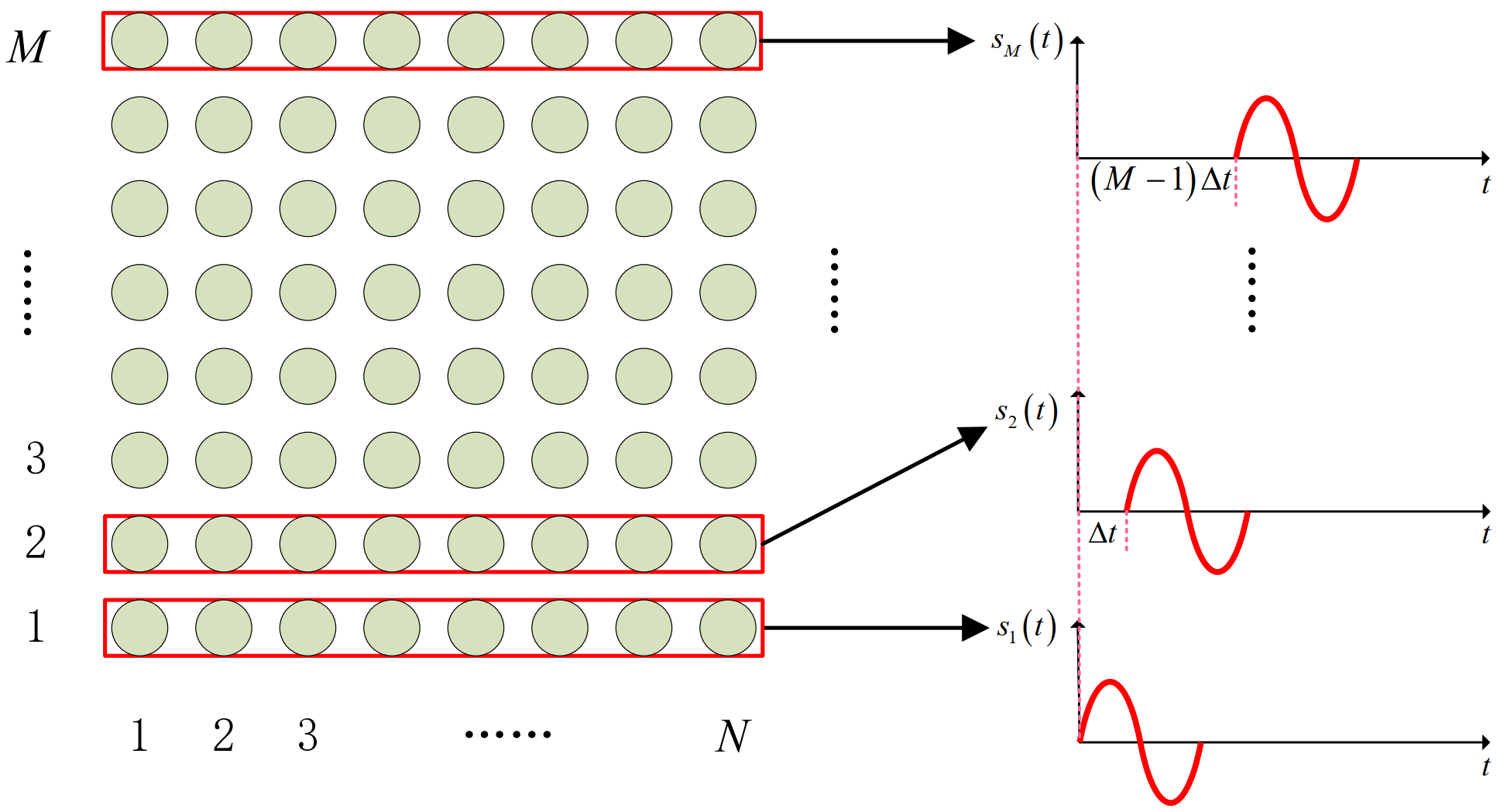}
    \caption{MIMO-STCA planar array signal model}
    \label{fig1}
\end{figure}

In this paper, we propose the design of MIMO-STCA radar, which introduces the transmitted time shift among transmit elements in different rows. Consider the planar array of MIMO-STCA radar is composed of M rows and N columns of elements. All elements are uniformly distributed with half-wavelength spacing between adjacent elements. The signal model is shown in Fig. \ref{fig1}. The transmitted time delay for each element in the $mth$ row can be represented as
\begin{small}
\begin{equation}
\tau_{m}=(m-1) \Delta t
\label{Equ1}
\end{equation}
\end{small}where $\Delta t$ is the time shift among transmit array elements in different rows.

\begin{figure}[!t]
\centering
\includegraphics[width=2.5in]{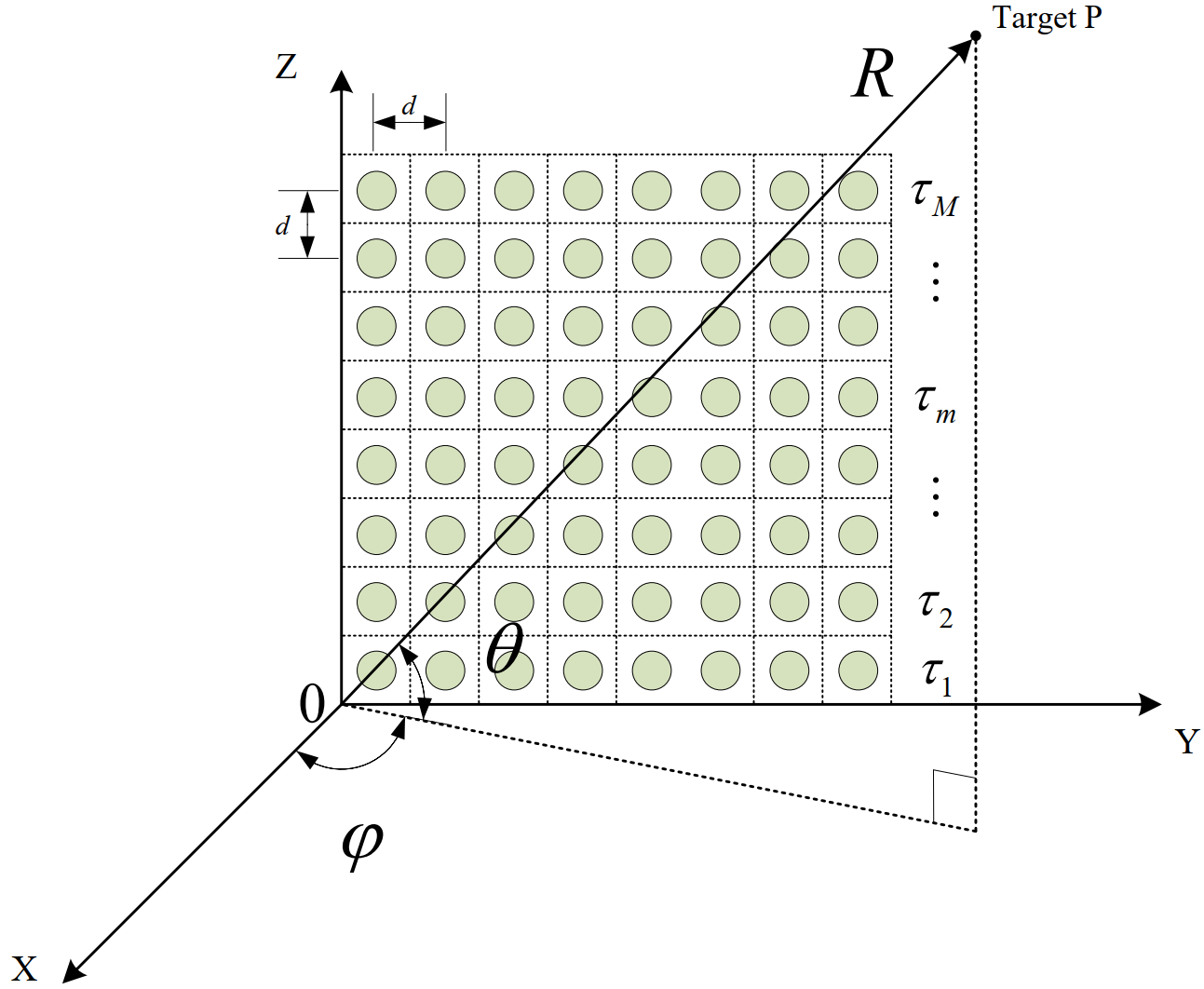}
\caption{The two-dimensional coordinate system for MIMO-STCA planar array}
\label{fig2}
\end{figure}

The MIMO-STCA radar antenna planar array is vertically placed on a horizontal plane, establishing the antenna planar array coordinate system as depicted in Fig. \ref{fig2}. In this figure, $\varphi$ and $\theta$ are azimuth angle and elevation angle, respectively. $R$ is the range between target and the radar array.

Assume the transmitted signal is a narrow-band signal. The orthogonal signal $\varphi_{m}(t)$ can be represented as:
\begin{small}
\begin{equation}
\varphi_{m}(t)=c_{m} g(t)
\label{Equ2}
\end{equation}
\end{small}where $c_{m}$ is the orthogonal coded signal, $g(t)$ is the linear frequency modulated (LFM) signal.
\begin{small}
\begin{equation}
\begin{array}{c} \int_{T} \varphi_{m}(t) \varphi_{m^{\prime}}(t) d t=0, \forall m, m^{\prime}=1,2, \ldots, M, m \neq m^{\prime}\end{array}
\label{Equ3}
\end{equation}
\end{small}
\begin{small}
\begin{equation}
g(t)=\operatorname{rect}\left(\frac{t}{T_{p}}\right) \exp \left(j \pi \mu t^{2}\right)
\label{Equ4}
\end{equation}
\end{small}where $T_p$ is the pulse duration, $\mu=B/T_p$ and $B$  is the bandwidth.

The signal transmitted by each element in the $mth$ row can be represented as
\begin{small}
\begin{equation}
\begin{array}{c}S_{m}(t)=c_{m} g(t-\tau_{m})exp(j2\pi f_0t)
\end{array}
\label{Equ5}
\end{equation}
\end{small}where $f_0$ is the carrier frequency.

Assume a far-field point target P located at azimuth angle $\varphi$, elevation angle $\theta$, and range $R$ from the antenna array.
Consider the origin element as the reference element, the unit vector in the direction of target P can be represented as $(u,v)$ ,where $u=cos\theta sin\varphi,v=sin\theta$. Therefore, the transmitted signal phase shift between the reference element and the element at the $mth$ row and $nth$ column is given by
\begin{small}
\begin{equation}
\begin{aligned}\alpha_{(m, n)}&=\frac{2 \pi}{\lambda}((n-1) d u+(m-1) d v)\\&=(n-1) \alpha_{y}+(m-1) \alpha_{z}
\end{aligned}
\label{Equ6}
\end{equation}
\end{small}where $\lambda=c/f_0$  is the  wavelength, $d$ represents the spacing between the adjacent elements. The superimposed transmitted signal from all MIMO-STCA radar elements can be given by
\begin{small}
    \begin{equation}
        \begin{aligned}S\big(t\big)&=\sum_{n=1}^{N}\sum_{m=1}^{M}e^{j\big(m-1\big)\alpha_{z}}e^{j\big(n-1\big)\alpha_{y}}s_{m}\big(t\big)\\&\approx e^{j2\pi f_{0}t}\sum_{n=1}^{N}\sum_{m=1}^{M}\varphi_{m}\big(t\big)e^{j\big(n-1\big)\alpha_{y}}e^{j\big(m-1\big)\alpha_{z}}e^{-j2\pi\mu\big(m-1\big)t\Delta t}\end{aligned}
        \label{Equ7}
    \end{equation}
\end{small}

Under the narrow-band assumption, the signal received by target P can be expressed as
\begin{small}
\begin{equation}
\begin{array}{c}\begin{aligned} S\left(t-\frac{\tau}{2}\right) & =e^{j 2 \pi f_{0}\left(t-\frac{\tau}{2}\right)} \sum_{n=1}^{N} \sum_{m=1}^{M} \varphi_{m}\left(t-\frac{\tau}{2}\right)  \\ &  e^{j(n-1) \alpha_{y}} e^{j(m-1) \alpha_{z}} e^{-j 2 \pi \mu(m-1)\left(t-\frac{R}{c}\right) \Delta t}\end{aligned}
\end{array}
\label{Equ8}
\end{equation}
\end{small}where $\tau=2R/c$ is the round-trip delay time between radar and target. Further, the radar echo received by the $m'th(m'=1,2, \ldots,M)$ row and $n'th(n'=1,2, \ldots,N)$ column element from target P’s reflection can be represented as
\begin{small}
\begin{equation}
\begin{aligned}x_{(m^{\prime},n^{\prime})}\begin{pmatrix}t\end{pmatrix}&=\xi_se^{j\left(n^{\prime}-1\right)\alpha_y}e^{j\left(m^{\prime}-1\right)\alpha_z}S\left(t-\tau\right)\end{aligned}
\label{Equ9}
\end{equation}
\end{small}where $\xi_{s}$ is the complex reflection coefficient of the target. The output of the received signal at the $mth$ row and $nth$ column element after down-conversion can be obtained as
\begin{small}
\begin{equation}
\begin{array}{c}\begin{aligned} \hat{x}_{\left(m^{\prime}, n^{\prime}\right)}(t)= & x_{\left(n^{\prime}, m^{\prime}\right)}(t) \cdot e^{-j 2 \pi f_{0} t} \end{aligned}
\end{array}
\label{Equ10}
\end{equation}
\end{small}

% The received signal of each element needs to undergo matched filtering. Beforehand, it is necessary to digitally mix the received signal from each receive element with $\Delta t$, which can be represented as
Before matched filtering, it is necessary to digitally mix the received signal from each receive element with $\Delta t$, which can be represented as
\begin{small}
\begin{equation}
\begin{array}{c}\begin{aligned} \tilde{x}_{\left(m^{\prime}, n^{\prime}\right)}(t)= & \hat{x}_{\left(m^{\prime}, n^{\prime}\right)}(t) \bullet e^{j 2 \pi \mu\left(m^{\prime}-1\right) t \Delta t} \\ = & \xi_{s} e^{-j 2 \pi f_{0} \tau} \frac{1-e^{j N \alpha_{y}}}{1-e^{j \alpha_{y}}} e^{j\left(n^{\prime}-1\right) \alpha_{y}} e^{j\left(m^{\prime}-1\right) \alpha_{z}}\\ & \varphi_{m^{\prime}}(t-\tau) e^{j\left(m^{\prime}-1\right) \alpha_{z}} e^{j 2 \pi \mu \frac{2 R}{c}\left(m^{\prime}-1\right) \Delta t} \\ & +\xi_{s} e^{-j 2 \pi f_{0} \tau} \frac{1-e^{j N \alpha_{y}}}{1-e^{j \alpha_{y}}} e^{j\left(n^{\prime}-1\right) \alpha_{y}} e^{j\left(m^{\prime}-1\right) \alpha_{z}} \\ & \times \sum_{m=1, m \neq m^{\prime}}^{M} \varphi_{m}(t-\tau) e^{j(m-1) \alpha_{z}} \\ &e^{-j 2 \pi \mu(m-1)\left(t-\frac{2 R}{c}\right) \Delta t} e^{j 2 \pi \mu\left(m^{\prime}-l\right)t \Delta t}\end{aligned}
\end{array}
\label{Equ11}
\end{equation}
\end{small}

Design matched filter functions for each receive element in the $m'th$  row as $h_{m^{\prime}}(t)=\varphi_{m^{\prime}}^{*}(-t)$. Moreover, since $\varphi_m\left(t\right)$ satisfies the orthogonality condition, the output of the matched filter can be obtained as

\begin{small}
\begin{equation}
\begin{array}{c}\begin{aligned} \bar{x}_{\left(m^{\prime}, n^{\prime}\right)}(t)&= \tilde{x}_{\left(m^{\prime}, n^{\prime}\right)}(t) * h_{m^{\prime}}(t) \\&=\beta_{s} e^{j\left(n^{\prime}-1\right) \alpha_{y}} e^{j 2\left(m^{\prime}-1\right) \alpha_{z}} e^{j 4 \pi \mu \frac{R}{c}\left(m^{\prime}-1\right) \Delta t} \end{aligned}
\end{array}
\label{Equ12}
\end{equation}
\end{small}where $*$ denotes the convolution operation, $\beta_{s}=\xi_{s} e^{-j 2 \pi f_{0} \tau} \frac{1-e^{j N \alpha_{y}}}{1-e^{j \alpha_{y}}}$ is the complex envelope of the received signal. Finally, the matched filter output for each element can be expressed as a vector, with the form as
% The output of the matched filter for the received signals of each element can ultimately be written as a vector, with the form as
\begin{small}
\begin{equation}
\begin{aligned}\mathbf{X}_s=&[\overline{x}_{(1,1)}\left(t\right),\overline{x}_{(1,2)}\left(t\right),\cdots,\overline{x}_{(1,N)}\left(t\right),\\ &\overline{x}_{(2,1)}\left(t\right),\overline{x}_{(2,2)}\left(t\right),\cdots,\overline{x}_{(M,N)}\left(t\right)]^T\\=&\beta_s\mathbf{a}_z\left(\theta,R\right)\otimes\mathbf{a}_y\left(\theta,\varphi\right)=\beta_s\mathbf{v}\left(\theta,\varphi,R\right)\end{aligned}
\label{Equ13}
\end{equation}
\end{small}where $\otimes$ denotes the Kronecker product, $\mathbf{v}(\theta, \varphi, R) \in \mathbb{C}^{M N \times 1}$, $\mathbf{a}_{y}(\theta, \varphi) \in \mathbb{C}^{N \times 1}$ and $\mathbf{a}_{z}(\theta, R) \in \mathbb{C}^{M \times 1}$ denote the array steering vector, the horizontal steering vector and vertical steering vector, respectively, which are expressed as 
% Specific expressions are given as follows
\begin{small}
\begin{equation}
\begin{aligned}\mathbf{a}_{y}\left(\theta,\varphi\right)&=\left[1,e^{j\alpha_{y}},\cdots,e^{j\left(N-1\right)\alpha_{y}}\right]^{T}\\&=\left[1,e^{j2\pi\frac{d}{\lambda}\cos\theta\sin\varphi},\cdots,e^{j2\pi\frac{d}{\lambda}\left(N-1\right)\mathrm{cos}\theta\sin\varphi}\right]^{T}\end{aligned}
\label{Equ14}
\end{equation}
\end{small}
\begin{small}
\begin{equation}
\begin{aligned}\mathbf{a}_z\left(\theta,R\right)&=\left[1,e^{j2\alpha_z}e^{j\cdot4\pi\mu\frac Rc\Delta t},\cdots,e^{j\:2\left(M-1\right)\alpha_z}e^{j\cdot4\pi\mu\frac Rc\left(M-1\right)\Delta t}\right]^T\\&=\left[1,e^{j\:4\pi\left(\frac d\lambda\sin\theta+\mu\frac Rc\Delta t\right)},\cdots,e^{j\:4\pi\left(M-1\right)\left(\frac d\lambda\sin\theta+\mu\frac Rc\Delta t\right)}\right]^T\end{aligned}
\label{Equ15}
\end{equation}
\end{small}

It can be observed that for MIMO-STCA radar, the horizontal steering vector is dependent solely on angle, while the vertical steering vector is a two-dimensional function of elevation angle and range. In contrast to traditional MIMO radar, it exhibits an extra degree of freedom in range domain. When the time shift in MIMO-STCA radar is set equal to 0, the vertical steering vector reduces to a function that depends only on the elevation angle. Therefore, its signal model is the same as that of the traditional MIMO radar.

Similarly, for a jammer with location $\left(\varphi_{j}, \theta_{j}, R_{j}\right)$ , which can generate MLJ, 
% after matched filtering, 
the jamming signal can be expressed as
\begin{equation}
\begin{array}{c}\mathbf{x}_{j}=\beta_{j} \mathbf{a}_{z}\left(\theta_{j}, R_{j}\right) \otimes \mathbf{a}_{y}\left(\theta_{j}, \varphi_{j}\right)=\beta_{j} \mathbf{v}\left(\theta_{j}, \varphi_{j}, R_{j}\right)
\end{array}
\label{Equ16}
\end{equation}
where $\beta_{j}$ is the complex envelope of the jamming signal. Therefore, the received signal can be expressed as
\begin{equation}
\mathbf{x}=\mathbf{x}_s+\mathbf{x}_j+\mathbf{x}_n
\label{Equ17}
\end{equation}
where $\mathbf{x}_{n}$ is the noise component, which can be assumed to be zero-mean white Gaussian distributed.

\section{Four-channel Adaptive beamforming algorithm for MLJ suppression}
\label{section:3}

In this section, we propose a four-channel adaptive beamforming algorithm, which introduces delta-delta beam as the auxiliary channel and utilizes sum-and-difference beamforming for adaptive MLJ cancellation. The key idea is suppressing MLJ with null along one direction while keeping the sum and difference beampatterns undistorted along another direction.

\subsection{Azimuth-elevation two-dimensional signal processing}
\begin{figure}[htpb]
\centering
\includegraphics[width=\columnwidth]{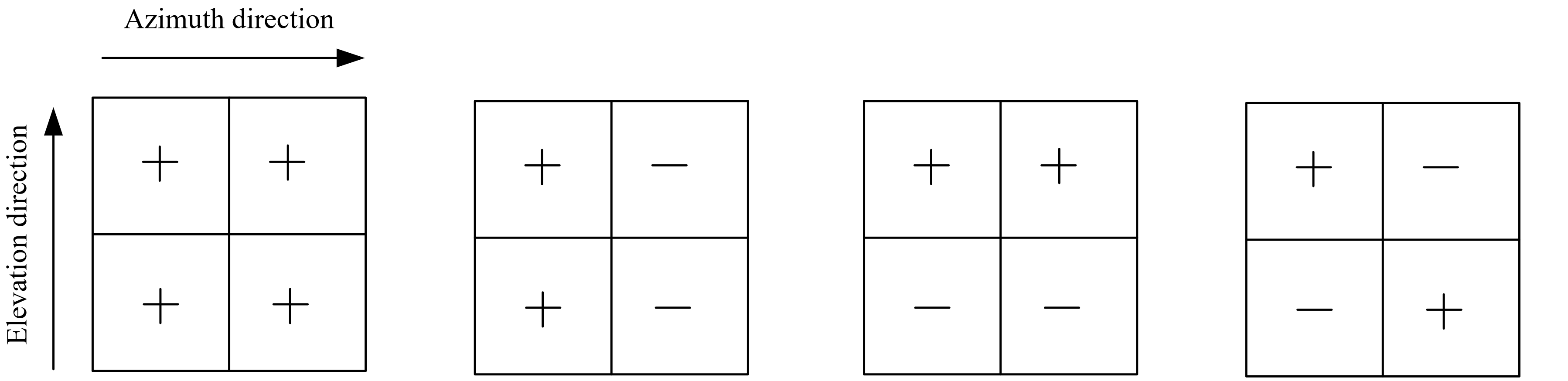}
\caption{Azimuth-elevation four-channel beamforming diagram. (a) Sum beam. (b) Delta-azimuth beam. (c) Delta-elevation beam. (d) Delta-delta beam}
\label{fig3}
\end{figure}
First, set $\Delta t$ equal to 0, thus the vertical steering vector changes to $\mathbf{a}_{ze}(\theta)=\left[1, e^{j 4 \pi \frac{d}{\lambda} \sin \theta}, \cdots, e^{j 4 \pi(M-1) \frac{d}{2} \sin \theta}\right]^{T}$ ,which is only dependent on the elevation angle, thus the array steering vector changes to $\mathbf{v}(\theta, \varphi)=\mathbf{a}_{ze}(\theta) \otimes \mathbf{a}_{y}(\theta, \varphi)$. As a result, the planar array becomes an azimuth-elevation two-dimensional planar array. Second, as shown in Fig. \ref{fig3}, the two-dimensional planar array is divided into four parts: upper-left, lower-left, upper-right, and lower-right, with the digital beam output denoted as $f_{1}, f_{2}, f_{3}$ and $ f_{4}$, respectively. We can obtain the quiescent four-channel beams named two-dimensional sum beam $f_{\Sigma}$ , delta-azimuth beam $f_{\Delta_{A}}$ , delta-elevation beam $f_{\Delta_{E}}$ , and delta-delta beam $f_{\Delta_{\Delta}}$ as
\begin{small}
\begin{equation}
\begin{array}{c}\left\{\begin{array}{l}f_{\Sigma}=\left(f_{1}+f_{2}\right)+\left(f_{3}+f_{4}\right) \\ f_{\Delta_{A}}=\left(f_{1}+f_{2}\right)-\left(f_{3}+f_{4}\right) \\ f_{\Delta_{E}}=\left(f_{1}+f_{3}\right)-\left(f_{2}+f_{4}\right) \\ f_{\Delta_{\Delta}}=\left(f_{1}+f_{4}\right)-\left(f_{2}+f_{3}\right)\end{array}\right.
\end{array}
\label{Equ18}
\end{equation}
\end{small}

Suppose the MIMO-STCA radar beam is directed to $\left(\varphi_{0}, \theta_{0}, R_{0}\right)$ and the target location is $\left(\varphi_{s}, \theta_{s}, R_{s}\right)$, the received signals of each element are weighted by $\mathbf{W}_{\Sigma-\Sigma}=\mathbf{a}_{ze}\left(\theta_{0}\right) \otimes \mathbf{a}_{y}\left(\theta_{0}, \varphi_{0}\right)$ , thus we can obtain the two-dimensional quiescent sum beampattern as
\begin{small}
\begin{equation}
\begin{aligned}g_\Sigma\left(u,v\right)&=\mathbf{W}_{\Sigma-\Sigma}^H\mathbf{v}\left(\theta,\varphi\right)\\&=\sum_{m=1}^M\sum_{n=1}^Ne^{j\left[\left(n-1\right)\left(\alpha_y-\alpha_{y0}\right)+2\left(m-1\right)\left(\alpha_z-\alpha_{z0}\right)\right]}\\&=\sum_{n=1}^Ne^{j\left(n-1\right)\left(\alpha_y-\alpha_{y0}\right)}\times\sum_{m=1}^Me^{j2\left(m-1\right)\left(\alpha_z-\alpha_{z0}\right)}\\&=g_{\sum_a}\left(u\right)\times g_{\sum_e}\left(v\right)\end{aligned}
\label{Equ19}
\end{equation}
\end{small}
% Decompose the double summation into the product of two summations, which can be written as
% \begin{small}
% \begin{equation}
% \begin{aligned}g_\Sigma\left(u,v\right)&=\sum_{n=1}^Ne^{j\left(n-1\right)\left(\alpha_y-\alpha_{y0}\right)}\times\sum_{m=1}^Me^{j2\left(m-1\right)\left(\alpha_z-\alpha_{z0}\right)}\\&=g_{\sum_a}\left(u\right)\times g_{\sum_e}\left(v\right)\end{aligned}
% \end{equation}
% \label{Equ20}
% \end{small}
% \begin{small}
% \begin{subequations}
% \begin{equation}
% \begin{aligned}g_\Sigma\left(u,v\right)&=\sum_{n=1}^Ne^{j\left(n-1\right)\left(\alpha_y-\alpha_{y0}\right)}\times\sum_{m=1}^Me^{j2\left(m-1\right)\left(\alpha_z-\alpha_{z0}\right)}\\&=g_{\sum_a}\left(u\right)\times g_{\sum_e}\left(v\right)\end{aligned}
% \end{equation}
% \label{Equ20}
% \tag{20}
% \end{subequations}  
% \end{small}
This indicates that the beampattern of a two-dimensional rectangular array is the product of independent row and column beampattern. This independence allows MLJ cancellation along one direction while keeping the sum and difference beampatterns undistorted along another direction.
% It is precisely because of their independence that we can cancel MLJ with null along one direction while keeping the sum and difference beampatterns undistorted along another direction.

% Similarly, the other three two-dimensional beampatterns can be separated as
% \begin{subequations}
% \begin{equation}
% g_{\Delta_{A}}(u, v)=g_{\Delta_{a}}(u) \times g_{\Sigma_{e}}(v)
% \tag{20b}
% \end{equation}
% \begin{equation}
% g_{\Delta_{E}}(u, v)=g_{\Sigma_{a}}(u) \times g_{\Delta_{e}}(v)
% \tag{20c}
% \end{equation}
% \begin{equation}
% g_{\Delta_{\Delta}}(u, v)=g_{\Delta_{a}}(u) \times g_{\Delta_{e}}(v)  
% \tag{20d}
% \end{equation}
% \end{subequations}

Moreover, (\ref{Equ18}) can be rewritten as
\begin{small}
\begin{equation}
\begin{array}{c}\left\{\begin{array}{l}
f_{\Sigma}=\beta_{s}g_{\Sigma}\left(u_{s}, v_{s}\right)+\beta_{j}g_{\Sigma}\left(u_{j}, v_{j}\right)+n_{\Sigma} \\ f_{\Delta_{A}}=\beta_{s} g_{\Delta_{A}}\left(u_{s}, v_{s}\right)+\beta_{j} g_{\Delta_{A}}\left(u_{j}, v_{j}\right)+n_{\Delta_{A}} \\ f_{\Delta_{E}}=\beta_{s} g_{\Delta_{E}}\left(u_{s}, v_{s}\right)+\beta_{j} g_{\Delta_{E}}\left(u_{j}, v_{j}\right)+n_{\Delta_{E}} \\ f_{\Delta_{\Delta}}=\beta_{s} g_{\Delta_{\Delta}}\left(u_{s}, v_{s}\right)+\beta_{j} g_{\Delta_{\Delta}}\left(u_{j}, v_{j}\right)+n_{\Delta_{\Delta}}\end{array}\right.
\end{array}
\label{Equ20}
\end{equation}
\end{small}where $n$ is the white Gaussian noise. As shown in (\ref{equ21}), using the sum beam as the main beam and the delta-elevation beam as the auxiliary beam for adaptive jamming cancellation, we obtain the adaptive azimuth-sum beam output $\hat{f}_{\Sigma_{A}}$. Similarly, we can obtain the adaptive azimuth-difference beam output $\hat{f}_{\Delta_{A}}$, the adaptive elevation-sum beam output $\hat{f}_{\Sigma_{E}}$ and the adaptive elevation-difference beam output $\hat{f}_{\Delta_{E}}$.

\begin{small}
\begin{subequations}
\begin{equation}
\begin{cases}\hat{f}_{\Sigma_A}=f_\Sigma-w_{a1}f_{\Delta_E}\\\hat{f}_{\Delta_A}=f_{\Delta_A}-w_{a2}f_{\Delta_\Delta}\end{cases}
\end{equation}
\begin{equation}
\begin{cases}\hat{f}_{\Sigma_E}=f_\Sigma-w_{e1}f_{\Delta_A}\\\hat{f}_{\Delta_E}=f_{\Delta_\mathrm{E}}-w_{e2}f_{\Delta_\Delta}\end{cases}
\end{equation}
\label{equ21}
\end{subequations}
\end{small}where $w_{a 1},w_{a 2},w_{e 1}$ and $w_{e 2}$ denote the adaptive weights, which are estimated from the sample values of the quiescent four-channel beams as
\begin{small}
\begin{subequations}
\begin{equation}
w_{a1}=\frac{E\left[f_{\Sigma} f_{\Delta_{E}}^{*}\right]}{E\left[f_{\Delta_{E}} f_{\Delta_{E}}^{*}\right]},w_{a2}=\frac{E\left[f_{\Delta_{A}} f_{\Delta_{\Delta}}^{*}\right]}{E\left[f_{\Delta_{\Delta}} f_{\Delta_{\Delta}}^{*}\right]}
\tag{22a}
\end{equation}
\begin{equation}
w_{e1}=\frac{E\left[f_{\Sigma} f_{\Delta_{A}}^{*}\right]}{E\left[f_{\Delta_{A}} f_{\Delta_{A}}^{*}\right]},w_{e2}=\frac{E\left[f_{\Delta_{E}} f_{\Delta_{\Delta}}^{*}\right]}{E\left[f_{\Delta_{\Delta}} f_{\Delta_{\Delta}}^{*}\right]}
\tag{22b}
\end{equation}
\end{subequations}
\end{small}

According to the correlation theory, we can obtain
\begin{small}
\begin{subequations}
\begin{equation}
E[f_{\Sigma}f_{\Delta_{E}}^{*}]=P_{j} g_{\Sigma}(u_{j},v_{j})g_{\Delta_{E}}^{*}(u_{j},v_{j})+P_{s} g_{\Sigma}(u_{s},v_{s})g_{\Delta_{E}}^{*}(u_{s},v_{s})
\tag{23a}
\end{equation}
\begin{equation}
E[f_{\Delta_{E}}f_{\Delta_{E}}^{*}]=P_{j} g_{\Delta_{E}}(u_{j},v_{j})g_{\Delta_{E}}^{*}(u_{j},v_{j})+P_{s} g_{\Delta_{E}}(u_{s},v_{s})g_{\Delta_{E}}^{*}(u_{s},v_{s}) + P_{n}
\tag{23b}
\end{equation}
\begin{equation}
E[f_{\Delta_{A}}f_{\Delta_{\Delta}}^{*}]=P_{j} g_{\Delta_{A}}(u_{j},v_{j})g_{\Delta_{\Delta}}^{*}(u_{j},v_{j})+P_{s} g_{\Delta_{A}}(u_{s},v_{s})g_{\Delta_{\Delta}}^{*}(u_{s},v_{s})
\tag{23c}
\end{equation}
\begin{equation}
E[f_{\Delta_{\Delta}}f_{\Delta_{\Delta}}^{*}]=P_{j} g_{\Delta_{\Delta}}(u_{j},v_{j})g_{\Delta_{\Delta}}^{*}(u_{j},v_{j})+P_{s} g_{\Delta_{\Delta}}(u_{s},v_{s})g_{\Delta_{\Delta}}^{*}(u_{s},v_{s})+P_{n}
\tag{23d}
\end{equation}
\begin{equation}
E[f_{\Sigma}f_{\Delta_{A}}^{*}]=P_{j} g_{\Sigma}(u_{j},v_{j})g_{\Delta_{A}}^{*}(u_{j},v_{j})+P_{s} g_{\Sigma}(u_{s},v_{s})g_{\Delta_{A}}^{*}(u_{s},v_{s})
\tag{23e}
\end{equation}
\begin{equation}
E[f_{\Delta_{A}}f_{\Delta_{A}}^{*}]=P_{j} g_{\Delta_{A}}(u_{j},v_{j})g_{\Delta_{A}}^{*}(u_{j},v_{j})+P_{s} g_{\Delta_{A}}(u_{s},v_{s})g_{\Delta_{A}}^{*}(u_{s},v_{s}) + P_{n}
\tag{23f}
\end{equation}
\begin{equation}
E[f_{\Delta_{E}}f_{\Delta_{\Delta}}^{*}]=P_{j} g_{\Delta_{E}}(u_{j},v_{j})g_{\Delta_{\Delta}}^{*}(u_{j},v_{j})+P_{s} g_{\Delta_{E}}(u_{s},v_{s})g_{\Delta_{\Delta}}^{*}(u_{s},v_{s})
\tag{23g}
\end{equation}
\end{subequations}
\end{small}where $P_{s},P_{j}$ and $P_{n}$ are the power of the target echo, MLJ and noise, respectively. Because $P_{j} \gg P_{s}, P_{j} \gg P_{n}$, we obtain
\begin{small}
\begin{subequations}
\begin{equation}
\left\{\begin{array}{l}w_{a1}=\frac{E\left[f_{\Sigma} f_{\Delta_{E}}^{*}\right]}{E\left[f_{\Delta_{E}} f_{\Delta_{E}}^{*}\right]} \approx \frac{g_{\Sigma}\left(u_{j}, v_{j}\right)}{g_{\Delta_{E}}\left(u_{j}, v_{j}\right)}=\frac{g_{\Sigma_{e}}\left(v_{j}\right)}{g_{\Delta_{e}}\left(v_{j}\right)} \\ w_{a 2}=\frac{E\left[f_{\Delta_{A}} f_{\Delta_{\Delta}}^{*}\right]}{E\left[f_{\Delta_{\Delta}} f_{\Delta_{\Delta}}^{*}\right]} \approx \frac{g_{\Delta_{A}}\left(u_{j}, v_{j}\right)}{g_{\Delta_{\Delta}}\left(u_{j}, v_{j}\right)}=\frac{g_{\Sigma_{e}}\left(v_{j}\right)}{g_{\Delta_{e}}\left(v_{j}\right)}\end{array}\right.
\end{equation}
\begin{equation}
\left\{\begin{array}{l}w_{e 1}=\frac{E\left[f_{\Sigma} f_{\Delta_{A}}^{*}\right]}{E\left[f_{\Delta_{A}} f_{\Delta_{A}}^{*}\right]} \approx \frac{g_{\Sigma}\left(u_{j}, v_{j}\right)}{g_{\Delta_{A}}\left(u_{j}, v_{j}\right)}=\frac{g_{\Sigma_{a}}\left(u_{j}\right)}{g_{\Delta_{a}}\left(u_{j}\right)} \\ w_{e 2}=\frac{E\left[f_{\Delta_{E}} f_{\Delta_{\Delta}}^{*}\right]}{E\left[f_{\Delta_{\Delta}} f_{\Delta_{\Delta}}^{*}\right]} \approx \frac{g_{\Delta_{E}}\left(u_{j}, v_{j}\right)}{g_{\Delta_{\Delta}}\left(u_{j}, v_{j}\right)}=\frac{g_{\Sigma_{a}}\left(u_{j}\right)}{g_{\Delta_{a}}\left(u_{j}\right)}\end{array}\right.
\end{equation}
\end{subequations}
\end{small}

It is obvious that $w_{a 1}=w_{a 2}=w_{a}, w_{e 1}=w_{e 2}=w_{e}$ . Thus, the adaptive four-channel beams can be rewritten as
\begin{small}
\begin{subequations}
\begin{equation}
\begin{cases}\hat{f}_{\sum_A}=\beta_s\left(g_\Sigma\left(u_s,v_s\right)-w_ag_{\Delta_E}\left(u_s,v_s\right)\right)+n_\Sigma-w_an_{\Delta_E}\\\hat{f}_{\Delta_A}=\beta_s\left(g_{\Delta_A}\left(u_s,v_s\right)-w_ag_{\Delta_\Delta}\left(u_s,v_s\right)\right)+n_{\Delta_A}-w_an_{\Delta_\Delta}\end{cases}
\end{equation}
\begin{equation}
\left\{\begin{array}{l}\hat{f}_{\sum_E}=\beta_s\left(g_\Sigma\left(u_s,v_s\right)-w_eg_{\Delta_A}\left(u_s,v_s\right)\right)+n_\Sigma-w_en_{\Delta_A}\\\\\hat{f}_{\Delta_E}=\beta_s\left(g_{\Delta_E}\left(u_s,v_s\right)-w_eg_{\Delta_\Delta}\left(u_s,v_s\right)\right)+n_{\Delta_E}-w_en_{\Delta_\Delta}\end{array}\right.
\end{equation}
\end{subequations}
\end{small}where the jamming is suppressed, and we can obtain the adaptive monopulse ratios in azimuth dimension and elevation dimension as
\begin{small}
\begin{subequations}
\begin{equation}
\hat{m}_{a}=\frac{\hat{f}_{\Delta_{A}}}{\hat{f}_{\Sigma_{A}}} \approx \frac{g_{\Delta_{A}}\left(u_{s}, v_{s}\right)-w_{a} g_{\Delta_{\Delta}}\left(u_{s}, v_{s}\right)}{g_{\Sigma}\left(u_{s}, v_{s}\right)-w_{a} g_{\Delta_{E}}\left(u_{s}, v_{s}\right)}=\frac{g_{\Delta_{a}}\left(u_{s}\right)}{g_{\Sigma_{a}}\left(u_{s}\right)}=m_{a}
\end{equation}
\begin{equation}
\hat{m}_{e}=\frac{\hat{f}_{\Delta_{e}}}{\hat{f}_{\Sigma_{e}}} \approx \frac{g_{\Delta_{E}}\left(u_{s}, v_{s}\right)-w_{e} g_{\Delta_{\Delta}}\left(u_{s}, v_{s}\right)}{g_{\Sigma}\left(u_{s}, v_{s}\right)-w_{e} g_{\Delta_{A}}\left(u_{s}, v_{s}\right)}=\frac{g_{\Delta_{e}}\left(v_{s}\right)}{g_{\Sigma_{e}}\left(v_{s}\right)}=m_{e}
\end{equation}
\label{Equ26}
\end{subequations}
\end{small}where ${m}_{a}$ and ${m}_{e}$ are the quiescent monopulse ratios in azimuth dimension and elevation dimension, respectively. (\ref{Equ26}) shows that the adaptive monopulse ratio along the azimuth (elevation) direction is maintained while canceling the MLJ along the elevation (azimuth) direction. In conclusion, the adaptive output ratio can still be utilized for angle estimation.

\setlength{\parskip}{0pt} % 设置段落间距

\subsection{Azimuth-range two-dimensional signal processing}
After the above process, setting $\Delta t=1 / B$, and we have determined the elevation angle $\theta$, so we can apply elevation angle dependence compensation to the vertical steering vector $\mathbf{a}_{z}(\theta, R)$, which can be represented as
\begin{small}
\begin{equation}
\begin{split}
\mathbf{a}_{zr}\left(R\right)&=\mathbf{a}_z\left(\theta,R\right)\odot\mathbf{c}\left(\theta\right) \\& =\left[1,e^{j4\pi\mu\frac Rc\Delta t},\cdotp\cdotp\cdotp,e^{j4\pi\left(M-1\right)\mu\frac Rc\Delta t}\right]^T
\end{split}
\label{Equ27}
\end{equation}
\end{small}where $\mathbf{c}(\theta)=\left[1, e^{-j 2 \pi \frac{d}{\lambda} \sin \theta}, \cdots, e^{-j 2 \pi \frac{d}{\lambda}(M-1) \sin \theta}\right]^{T}$, $\odot$ denotes the Hadamard product. After compensation, the vertical steering changes to $\mathbf{a}_{zr}(R)$, which is solely dependent on the range. Thus, the planar array becomes an azimuth-range two-dimensional planar array.

Similar to azimuth-elevation two-dimensional planar array,
% as shown in Fig. \ref{fig5},
we can obtain the output from upper-left part, lower-left part, upper-right part and lower-right part, denoting them as $r_{1}$, $r_{2}$, $r_{3}$ and $r_{4}$, respectively. And we can also obtain the quiescent four-channel beams as
\begin{small}
\begin{equation}
\begin{array}{c}\left\{\begin{array}{l}r_{\Sigma}=\left(r_{1}+r_{2}\right)+\left(r_{3}+r_{4}\right) \\ r_{\Delta_{A}}=\left(r_{1}+r_{2}\right)-\left(r_{3}+r_{4}\right) \\ r_{\Delta_{R}}=\left(r_{1}+r_{3}\right)-\left(r_{2}+r_{4}\right) \\ r_{\Delta_{\Delta}}=\left(r_{1}+r_{4}\right)-\left(r_{2}+r_{3}\right)\end{array}\right.
\end{array}
\label{Equ28}
\end{equation}
\end{small}

The received signals of each element are weighted by $\mathbf{W}_{\Sigma-\Sigma}^{\prime}=\mathbf{a}_{zr}\left(R_{0}\right) \otimes \mathbf{a}_{y}\left(\theta_{0}, \varphi_{0}\right)$, thus the azimuth-range two-dimensional quiescent sum beampattern is
\begin{small}
\begin{equation}
\begin{aligned}p_\Sigma\left(u,R\right)&=\mathbf{W}_{\Sigma-\Sigma}^{H}\left[\mathbf{a}_{zr}\left(R\right)\otimes\mathbf{a}_{y}\left(\theta,\varphi\right)\right]\\&=\sum_{m=1}^M\sum_{n=1}^N\mathrm{e}^{j\left(\left(n-1\right)\left(\alpha_y-\alpha_{y0}\right)+4\left(m-1\right)\pi\mu\frac{\left(R-R_0\right)}{c}\Delta t\right)}\\&=\sum_{n=1}^N\mathrm{e}^{j\left(\left(n-1\right)\left(\alpha_y-\alpha_{y0}\right)\right)}\times\sum_{m=1}^M\mathrm{e}^{j4\left(m-1\right)\pi\mu\frac{\left(R-R_0\right)}c\Delta t}\\&=p_{\Sigma_a}\left(u\right)\times p_{\Sigma_r}\left(R\right)\end{aligned}
\label{Equ29}
\end{equation}
\end{small}

Then, we can rewrite the four-channel beams as
\begin{small}
\begin{equation}
\begin{array}{c}\left\{\begin{array}{l}
r_{\Sigma}=\beta_{s} p_{\Sigma}\left(u_{s}, R_{s}\right)+\beta_{j} p_{\Sigma}\left(u_{j}, R_{j}\right)+n_{\Sigma} \\ r_{\Delta_{A}}=\beta_{s} p_{\Delta_{A}}\left(u_{s}, R_{s}\right)+\beta_{j} p_{\Delta_{A}}\left(u_{j}, R_{j}\right)+n_{\Delta_{A}} \\ r_{\Delta_{R}}=\beta_{s} p_{\Delta_{R}}\left(u_{s}, R_{s}\right)+\beta_{j} p_{\Delta_{R}}\left(u_{j}, R_{j}\right)+n_{\Delta_{R}} \\ r_{\Delta_{\Delta}}=\beta_{s} p_{\Delta_{\Delta}}\left(u_{s}, R_{s}\right)+\beta_{j} p_{\Delta_{\Delta}}\left(u_{j}, R_{j}\right)+n_{\Delta_{\Delta}}\end{array}\right.
\end{array}
\label{Equ30}
\end{equation}
\end{small}

Similar to azimuth-elevation two-dimensional planar array, the adaptive range sum and difference beams can be obtained as
\begin{small}
\begin{equation}
\begin{cases}\hat{r}_{\sum_R}=r_\Sigma-w_{r1}r_{\Delta_A}\\\hat{r}_{\Delta_R}=r_{\Delta_R}-w_{r2}r_{\Delta_\Delta}&\end{cases}
\label{Equ31}
\end{equation}
\end{small}where $w_{r 1}$ and $w_{r 2}$ denote the adaptive weights, which can be obtained as

\begin{equation}
{w}_{r1}=\frac{E{\left[r_{\Sigma}r_{\Delta_{A}}^{*}\right]}}{E{\left[r_{\Delta_{A}}r_{\Delta_{A}}^{*}\right]}},w_{r2}=\frac{E{\left[r_{\Delta_{R}}r_{\Delta_\Delta}^*\right]}}{E{\left[r_{\Delta_{\Delta}}r_{\Delta_{\Delta}}^{*}\right]}}
\label{Equ32}
\end{equation}

According to the correlation theory, we can obtain
\begin{small}
\begin{subequations}
\begin{equation}
E\left[r_\Sigma r_{\Delta_\mathrm{A}}^*\right] = P_s p_\Sigma(u_s, R_s) p_{\Delta_A}^*(u_s, R_s) + P_j p_\Sigma(u_j, R_j) p_{\Delta_A}^*(u_j, R_j)
\tag{33a}
\end{equation}
\begin{equation}
E\left[r_{\Delta_A} r_{\Delta_A}^*\right] = P_s p_{\Delta_A}(u_s, R_s) p_{\Delta_A}^*(u_s, R_s) + P_j p_{\Delta_A}(u_j, R_j) p_{\Delta_A}^*(u_j, R_j) + P_n 
\tag{33b}
\end{equation}
\begin{equation}
E\left[r_{\Delta_R} r^*_{\Delta_{\Delta}}\right] = P_s p_{\Delta_R}(u_s, R_s) p_{\Delta_\Delta}^*(u_s, R_s) + P_j p_{\Delta_R}(u_j, R_j) p_{\Delta_\Delta}^*(u_j, R_j) 
\tag{33c}
\end{equation}
\begin{equation}
E\left[r_{\Delta_{\Delta}} r^*_{\Delta_{\Delta}}\right] = P_s p_{\Delta_{\Delta}}(u_s, R_s) p_{\Delta_\Delta}^*(u_s, R_s) + P_j p_{\Delta_{\Delta}}(u_j, R_j) p_{\Delta_\Delta}^*(u_j, R_j) + P_n
\tag{33d}
\end{equation}
\end{subequations}
\end{small}

Similarly, since $P_{j} \gg P_{s}, P_{j} \gg P_{n}$, we can obtain

\begin{equation}
\begin{array}{c}\left\{\begin{array}{l}w_{r 1}=\frac{E\left[r_{\Sigma} r^{*}{ }_{\Delta_{A}}\right]}{E\left[r_{\Delta_{A}} r^{*}{ }_{\Delta_{A}}\right]} \approx \frac{p_{\Sigma}\left(u_{j}, R_{j}\right)}{p_{\Delta_{A}}\left(u_{j}, R_{j}\right)}=\frac{p_{\Sigma_{a}}\left(u_{j}\right)}{p_{\Delta_{a}}\left(u_{j}\right)} \\ w_{r 2}=\frac{E\left[r_{\Delta_{R}} r^{*}{ }_{\Delta_{\Delta}}\right]}{E\left[r_{\Delta_{\Delta}} r^{*}{ }_{\Delta_{\Delta}}\right]} \approx \frac{p_{\Delta_{R}}\left(u_{j}, R_{j}\right)}{p_{\Delta_{\Delta}}\left(u_{j}, R_{j}\right)}=\frac{p_{\Sigma_{a}}\left(u_{j}\right)}{p_{\Delta_{a}}\left(u_{j}\right)}\end{array}\right.
\end{array}
\label{Equ34}
\end{equation}

It can be concluded that $w_{r 1}=w_{r 2}=w_{r}$, so we can obtain the adaptive range sum and difference beams as
\begin{small}
\begin{equation}
\begin{array}{c}\left\{\begin{array}{l}\hat{r}_{\Sigma_{R}}=\beta_{s}\left(p_{\Sigma}\left(u_{s}, R_{s}\right)-w_{r} p_{\Delta_{A}}\left(u_{s}, R_{s}\right)\right)+n_{\Sigma}-w_{r} n_{\Delta_{A}} \\ \hat{r}_{\Delta_{R}}=\beta_{s}\left(p_{\Delta_{R}}\left(u_{s}, R_{s}\right)-w_{r} p_{\Delta_{\Delta}}\left(u_{s}, R_{s}\right)\right)+n_{\Delta_{R}}-w_{r} n_{\Delta_{\Delta}}\end{array}\right.
\end{array}
\label{Equ37}
\end{equation}
\end{small}

Further, the ratio of the adaptive range difference beam to the sum beam is
\begin{equation}
\begin{array}{c}\hat{m}_{R}=\frac{\hat{r}_{\Delta_{R}}}{\hat{r}_{\Sigma_{R}}} \approx \frac{p_{\Delta_{R}}\left(u_{s}, R_{s}\right)-w_{r} p_{\Delta_{\Delta}}\left(u_{s}, R_{s}\right)}{p_{\Sigma}\left(u_{s}, R_{s}\right)-w_{r} p_{\Delta_{A}}\left(u_{s}, R_{s}\right)}=\frac{p_{\Delta_{r}}\left(R_{s}\right)}{p_{\Sigma_{r}}\left(R_{s}\right)}=m_{R}
\end{array}
\label{Equ36}
\end{equation}
where $m_{R}$ is the quiescent monopulse ratio in the range dimension. This equation indicates that the extra DOF in range dimension, which is introduced by MIMO-STCA radar can be used to estimate the target range. Furthermore, by maintaining the adaptive monopulse ratio in range dimension, ensuring the high performance of range estimation.

\section{Row-column Adaptive Beamforming Algorithm for Joint Suppression of Multiple MLJs and SLJs}
\label{section:4}

% Regarding the issue that the four-channel adaptive beamforming algorithm exploits 2 DOFs in both the horizontal and vertical directions, thus it can only effectively suppress one MLJ at most. Further, in this section, the row-column adaptive beamforming algorithm is proposed for MIMO-STCA radar to estimate the target angle and range with multiple MLJs and SLJs. This algorithm first conducts adaptive jammings cancellation for each row or column, followed by monopulse sum and difference beamforming. By treating each row or column as an adaptive subarray, multiple DOFs can be utilized to suppress multiple jammings simultaneously. Then, two-dimensional sum and difference beams can be obtained for angle and range estimation.
The four-channel adaptive beamforming algorithm uses 2 DOFs in both horizontal and vertical directions, limiting it to suppressing only one MLJ. To address this issue, we further propose the row-column adaptive beamforming algorithm for MIMO-STCA radar, enabling target angle and range joint estimation while suppressing multiple MLJs and SLJs. This algorithm first conducts adaptive jammings cancellation for each row or column. By treating each row or column as an adaptive subarray, multiple DOFs can be utilized to suppress multiple jammings simultaneously. Then, two-dimensional sum and difference beams can be obtained for parameters estimation.

\subsection{Azimuth-elevation two-dimensional signal processing}

With multiple MLJs and SLJs, assuming the number of jammings is $N_{J}$, the target location is $\left(\theta_{s}, \varphi_{s}, R_{s}\right)$, the received signal from the MIMO-STCA array elements is formed as
\begin{small}
\begin{equation}
\begin{aligned}\mathbf{x}&=\mathbf{x}_s+\sum_{n_j=1}^{N_j}\mathbf{x}_{n_j}+\mathbf{x}_n\end{aligned}
\label{Equ37}
\end{equation}
\end{small}

\begin{figure}
\centering
\includegraphics[width=\columnwidth]{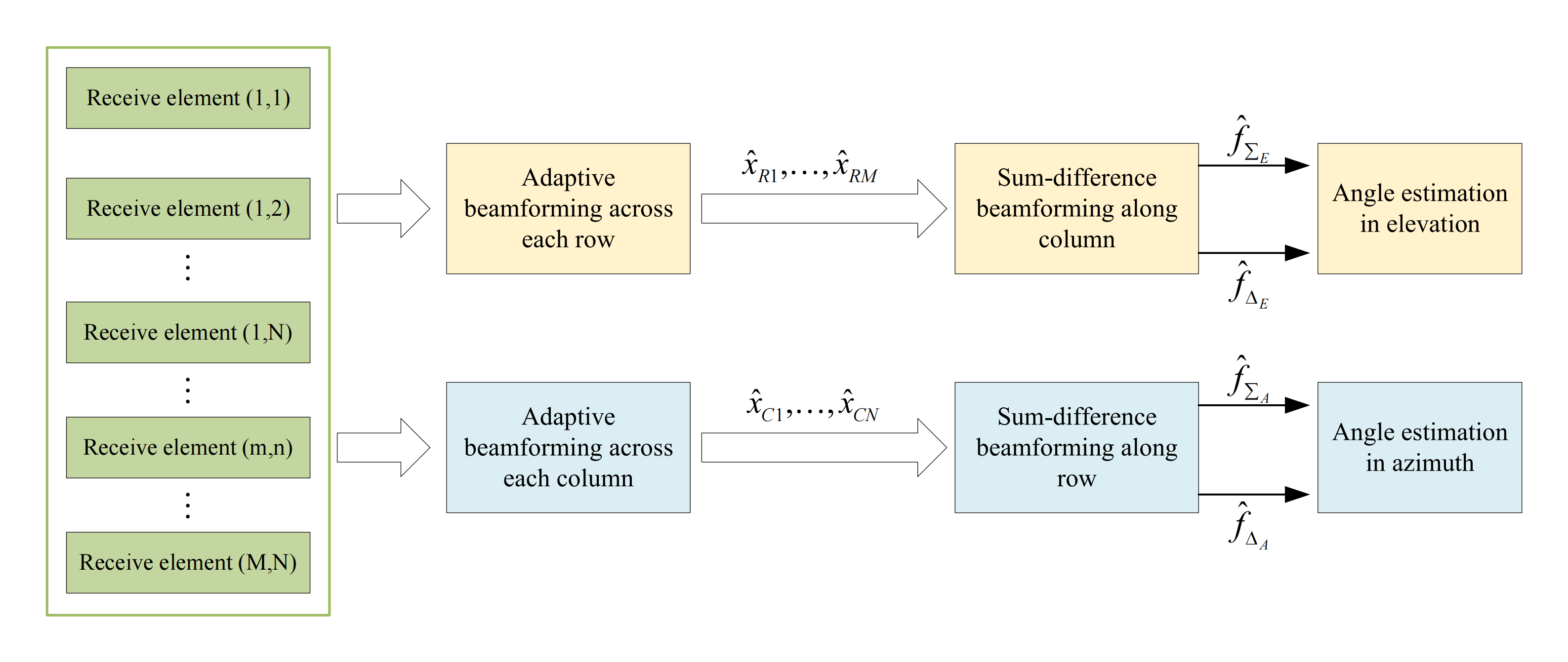}
\caption{Process of row-column adaptive beamforming algorithm}
\label{fig6}
\end{figure}

Fig. \ref{fig6} illustrates the row-column adaptive beamforming algorithm process. First, setting $\Delta t = 0$ transforms the planar array into an azimuth-elevation two-dimensional planar array, similar to the four-channel adaptive beamforming algorithm. Second, 1-D adaptive beamforming is performed across each row or column to cancel multiple jammings. Third, adaptive sum and difference beams are formed in the other dimension independently. Finally, monopulse angle estimation is applied.

\begin{figure}
\centering
\includegraphics[width=\columnwidth]{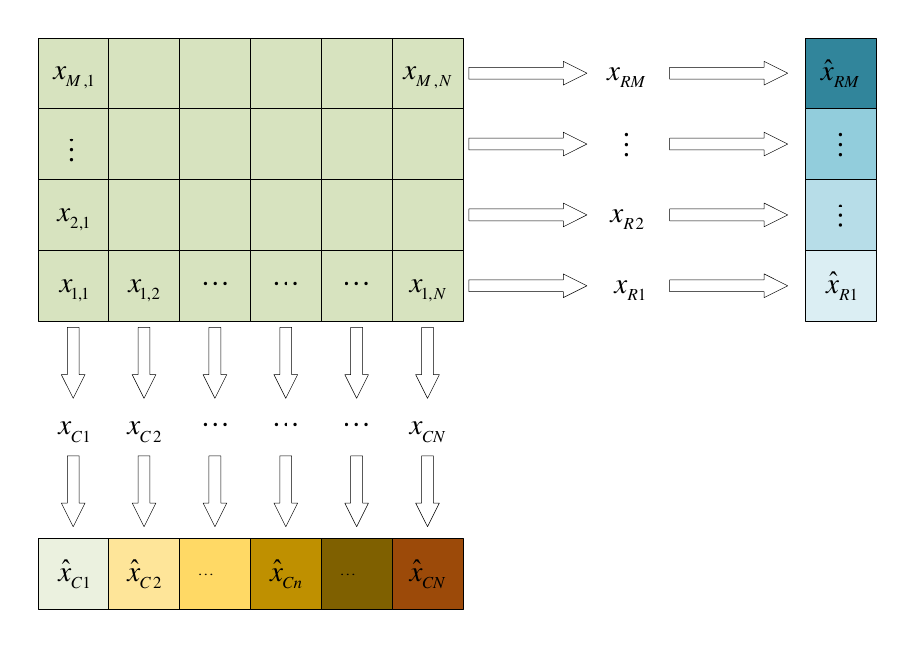}
\caption{Adaptive beamforming across rows and columns}
\label{fig7}
\end{figure}

The jammings cancellation via adaptive beamforming is accomplished across rows and columns separately, illustrating in Fig. \ref{fig7}. The received data vectors for the $mth$ row and the $nth$ column are given as follows
\begin{small}
\begin{subequations}
\begin{equation}
\mathbf{x}_{Rm}=\left[x_{m,1},x_{m,2},\ldots,x_{m,N}\right]^T,m=1,2,\ldots,M
\end{equation}
\begin{equation}
\mathbf{x}_{Cn}=\left[x_{1,n},x_{2,n},\ldots,x_{M,n}\right]^T,n=1,2,\ldots,N
\end{equation}
\end{subequations}
\end{small}

The adaptive beamforming in the $mth$ row is given as follows
\begin{small}
\begin{equation}
\hat{x}_{Rm}=x_{m,1}+\hat{w}_{m,2}^Rx_{m,2}+\ldots+\hat{w}_{m,N}^RX_{m,N}
\label{Equ39}
\end{equation}
\end{small}where ${w}^{R}$ is the adaptive weight. The adaptive weight vector of the $mth$ row can be written as
\begin{small}
\begin{equation}
\mathbf{\hat{W}}_m^R=\left[1,\hat{w}_{m,2}^R,\ldots,\hat{w}_{m,N}^R\right]^T
\label{Equ40}
\end{equation}
\end{small}

The weight is computed using the minimum variance distortionless response (MVDR) method, expressing as follows
\begin{small}
\begin{equation}
\begin{array}{c}\left\{\begin{array}{c}\min\left(\hat{\mathbf{W}}_{m}^{R}\right)^{H} \mathbf{R}_{X_{Rm}} \hat{\mathbf{W}}_{m}^{R} \\ \text {s.t.}\left(\hat{\mathbf{W}}_{m}^{R}\right)^{H} \mathbf{a}_{y}\left(\theta_{0}, \varphi_{0}\right)=1\end{array}\right.
\label{Equ41}
\end{array}
\end{equation}
\end{small}where $\mathbf{R}_{X_{Rm}}=E\left[\mathbf{x}_{Rm} \mathbf{x}_{Rm}^{H}\right]$ is the sample covariance matrix of the received data vector in the $mth$ row. The solution is derived using the Lagrange multiplier method, yielding the adaptive weight vector as
\begin{small}
\begin{equation}
\begin{cases}\hat{\mathbf{W}}_m^R={\alpha}(\mathbf{R}_{X_{Rm}})^{-1}\mathbf{a}_{y}\left({\theta}_0,\varphi_0\right)\\{\alpha}=1/\left(\mathbf{a}_{y}\left({\theta}_0,\varphi_0\right)\right)^H\left(\mathbf{R}_{X_{Rm}}\right)^{-1}\mathbf{a}_{y}\left({\theta}_0,\varphi_0\right)\end{cases}
\label{Equ42}
\end{equation}
\end{small}

The relationship between the steering vector of the $mth$ row and the $\left(m+1\right)th$ row can be represented as 
\begin{small}
\begin{equation}
\mathbf{v}_{m+1}\left(\theta,\varphi\right)=\mathrm{e}^{j4\pi\frac{d}{\lambda}\sin\theta}\times\mathbf{v}_m\left(\theta,\varphi\right)
\label{Equ43}
\end{equation}
\end{small}

Consequently, the relationship between the adaptive weight vector of the $mth$ row and the $(m+1)th$ row can be represented as
\begin{small}
\begin{equation}
\mathbf{\hat{W}}_{m+1}^R=\mathbf{e}^{j4\pi\frac{d}{\lambda}\sin\theta}\times\mathbf{\hat{W}}_m^R
\label{Equ44}
\end{equation}
\end{small}

The output of adaptive beam in the $mth$  row is given as
\begin{small}
\begin{equation}
\hat{x}_{Rm}=\left(\hat{\mathbf{W}}_m^R\right)^H\mathbf{x}_{Rm}
\label{Equ45}
\end{equation}
\end{small}

The adaptive $nth$ column beamforming in elevation dimension can be performed in a similar manner. The adaptive $nth$  column beamforming is given by the following
\begin{small}
\begin{equation}
\hat{x}_{Cn}=x_{1,n}+\hat{w}_{2,n}^Cx_{2,n}+\ldots+\hat{w}_{M,n}^Cx_{M,n}
\label{Equ46}
\end{equation}
\end{small}

The adaptive weight vector of the $nth$ column is given by
\begin{small}
\begin{equation}
\hat{\mathbf{W}}_n^C=\left[1,\hat{w}_{2,n}^C,\ldots,\hat{w}_{M,n}^C\right]^T
\label{Equ47}
\end{equation}
\end{small}

Similarly, $ \mathbf{W}_{n}^{c}$ can be obtained as follows
\begin{small}
\begin{equation}
\begin{cases}\hat{\mathbf{W}}_n^C={\beta}(\mathbf{R}_{X_{Cn}})^{-1}\mathbf{a}_{ze}\left({\theta}_0\right)\\{\beta}=1/\left(\mathbf{a}_{ze}\left({\theta}_0\right)\right)^H\left(\mathbf{R}_{X_{Cn}}\right)^{-1}\mathbf{a}_{ze}\left({\theta}_0\right)\end{cases}
\label{Equ48}
\end{equation}
\end{small}where $\mathbf{R}_{X_{c_{n}}}$ is the sample covariance matrix of the received data vector in the $nth$ column. The relationship between the adaptive weight vector of the $nth$ column and the $ \left(n+1\right)th$ column can be represented as
\begin{small}
\begin{equation}
\mathbf{\hat{W}}_{n+1}^C=e^{j2\pi\frac d\lambda\cos\theta\sin\varphi}\mathbf{\hat{W}}_n^C
\label{Equ49}
\end{equation}
\end{small}

The output of adaptive beam in the $nth$ column is given
\begin{small}
\begin{equation}
\hat{x}_{Cn}=\left(\hat{\mathbf{W}}_n^C\right)^H\mathbf{x}_{Cn}
\label{Equ50}
\end{equation}
\end{small}

\subsection{Azimuth-range two-dimensional signal processing}

Similarly, the elevation angle dependence compensation is applied to the vertical steering vector $\mathbf{a}_{z}(\theta, R)$, transforming the planar array into an azimuth-range two-dimensional planar array. The jammings cancellation is accomplished in azimuth dimension. The received data vectors for the $mth$ row are given as
\begin{small}
\begin{equation}
\mathbf{y}_{Rm}=\left[y_{m,1},y_{m,2},\ldots,y_{m,N}\right]^T,m=1,2,\ldots,M
\label{Equ51}
\end{equation}
\end{small}

The adaptive beamforming in the $mth$ row is given as
\begin{small}
\begin{equation}
\hat{y}_{Rm}=y_{m,1}+w_{m,2}^{\prime R}y_{m,2}+\ldots+w_{m,N}^{\prime R}y_{m,N}
\label{Equ52}
\end{equation}
\end{small}

The adaptive weight vector $\mathbf{W}_{m}^{\prime R}$ can be written as
\begin{small}
\begin{equation}
\mathbf{W}_m^{\prime R}=\left[1,w_{m,2}^{\prime R},\ldots,W_{m,N}^{\prime R}\right]^T
\label{Equ53}
\end{equation}
\end{small}

Similarly, $\mathbf{W}_{m}^{\prime R}$ can be derived as follows
\begin{small}
\begin{equation}
\begin{cases}\min\left(\mathbf{W}_m^{\prime R}\right)^H\mathbf{R}_{y_{Rm}}\mathbf{W}_m^{\prime R}\\s.t.\left(\mathbf{W}_m^{\prime R}\right)^H\mathbf{a}_y\left(\theta_0,\varphi_0\right)=1\end{cases}
\label{Equ54}
\end{equation}
\end{small}where $\mathbf{R}_{y_{Rm}}$ is the sample covariance matrix of the received data vector in the $mth$ row. Then, we can obtain
\begin{small}
\begin{equation}
\begin{cases}\mathbf{W}_m^{\prime R}={\alpha}(\mathbf{R}_{y_{Rm}})^{-1}\mathbf{a}_{y}\left({\theta}_0,\varphi_0\right)\\{\alpha}=1/\left(\mathbf{a}_{y}\left({\theta}_0,\varphi_0\right)\right)^H\left(\mathbf{R}_{y_{Rm}}\right)^{-1}\mathbf{a}_{y}\left({\theta}_0,\varphi_0\right)\end{cases}
\label{Equ55}
\end{equation}
\end{small}

The output of adaptive beam in the $mth$ row is given by
\begin{small}
\begin{equation}
\hat{y}_{Rm}=\left(\mathbf{W}_m^{\prime R}\right)^H\mathbf{y}_{Rm}
\label{Equ56}
\end{equation}
\end{small}

To sum up, all the M or N DOFs have been exploited in the adaptive row or column beamforming, which means theoretically up to $M-1$ or $N-1$ MLJs and/or SLJs can be adaptive suppressed across each row or each column.

\subsection{Adaptive monopulse parameters estimation}

After the adaptive processing, the three adaptive beams in three dimensions are given as follows
\begin{small}
\begin{subequations}
\begin{equation}
\hat{\mathbf{x}}_E=\left[\hat{x}_{R1},\hat{x}_{R2},\ldots,\hat{x}_{RM}\right]
\end{equation}
\begin{equation}
\hat{\mathbf{x}}_A=\begin{bmatrix}\hat{x}_{C1},\hat{x}_{C2},\ldots,\hat{x}_{CN}\end{bmatrix}
\end{equation}
\begin{equation}
\hat{\mathbf{x}}_R=\begin{bmatrix}\hat{y}_{R1},\hat{y}_{R2},\ldots,\hat{y}_{RM}\end{bmatrix}
\end{equation}
\end{subequations}
\end{small}where $\hat{x}_{E},\hat{x}_{A}$ and $\hat{x}_{R}$ represent the adaptive beams in elevation, azimuth and range dimensions, respectively. With adaptive column beam outputs, the sum-and-difference beams are formed along the row. Similarly, with adaptive row beam outputs, the sum-and-difference beams are formed along the column. Thus, the six monopulse beams are given as
\begin{small}
\begin{subequations}
\begin{equation}
\hat{f}_{\sum_E}=\left(\mathbf{w}_{\sum_E}\right)^H\hat{\mathbf{x}}_E\quad\hat{\boldsymbol{f}}_{\Delta_E}=\left(\mathbf{w}_{\Delta_E}\right)^H\mathbf{\hat{x}}_E
\end{equation}
\begin{equation}
\hat{f}_{\sum_A}=\left(\mathbf{w}_{\sum_A}\right)^H\hat{\mathbf{x}}_A\quad\hat{\boldsymbol{f}}_{\Delta_A}=\left(\mathbf{w}_{\Delta_A}\right)^H\hat{\mathbf{x}}_A
\end{equation}
\begin{equation}
\hat{f}_{\sum_R}=\left(\mathbf{w}_{\sum_R}\right)^H\hat{\mathbf{x}}_R\quad\hat{f}_{\Delta_R}=\left(\mathbf{w}_{\Delta_R}\right)^H\hat{\mathbf{x}}_R
\end{equation}
\end{subequations}
\end{small}where $\mathbf{w}_{\Sigma_{E}}$, $\mathbf{w}_{\Delta_{E}}$, $\mathbf{w}_{\Sigma_{A}}$, $\mathbf{w}_{\Delta_{A}}$, $\mathbf{w}_{\Sigma_{R}}$ and $\mathbf{w}_{\Delta_{R}}$ are the quiescent sum and difference weight vectors along elevation, azimuth, and range dimensions, respectively, as if monopulse beamforming is performed on a ULA. They are given by

\begin{small}
\begin{subequations}
\begin{equation}
\mathbf{w}_{\sum_E} = \mathbf{a}_{ze}\left(\theta_0\right)\left[1,e^{j4\pi\frac{d}{\lambda}\sin\theta_0},\cdots,e^{j4\pi(M-1)\frac{d}{\lambda}\sin\theta_0}\right]^T
\end{equation}
\begin{equation}
\mathbf{w}_{\Delta_E} = \mathbf{a}_{ze}\left(\theta_0\right) \odot \mathbf{c}_{\Delta M} = \left[1,e^{j4\pi\frac{d}{\lambda}\sin\theta_0},\cdots,-e^{j4\pi(M-1)\frac{d}{\lambda}\sin\theta_0}\right]^T
\end{equation}
\begin{equation}
\mathbf{w}_{\Sigma_A} = \mathbf{a}_y\left(\theta_0,\varphi_0\right) = \left[1,e^{j2\pi\frac{d}{\lambda}\cos\theta_0\sin\varphi_0},\cdots,e^{j2\pi\frac{d}{\lambda}(N-1)\cos\theta_0\sin\varphi_0}\right]^T
\end{equation}
\begin{equation}
\begin{aligned}
\mathbf{w}_{\Delta_A} &= \mathbf{a}_y\left(\theta_0,\varphi_0\right) \odot \mathbf{c}_{\Delta N} \\&= \left[1,e^{j2\pi\frac{d}{\lambda}\cos\theta_0\sin\varphi_0},\cdots,-e^{j2\pi\frac{d}{\lambda}(N-1)\cos\theta_0\sin\varphi_0}\right]^T
\end{aligned}
\end{equation}
\begin{equation}
\mathbf{w}_{\sum_R} = \mathbf{a}_{zr}\left(R_0\right)= \left[1,e^{j4\pi\mu\frac{R_0}{c}\Delta t},\cdots,e^{j4\pi(M-1)\mu\frac{R_0}{c}\Delta t}\right]^T 
\end{equation}
\begin{equation}
\mathbf{w}_{\Delta_R} = \mathbf{a}_{zr}\left(R_0\right) \odot \mathbf{c}_{\Delta M} = \left[1,e^{j4\pi\mu\frac{R_0}{c}\Delta t},\cdots,-e^{j4\pi(M-1)\mu\frac{R_0}{c}\Delta t}\right]^T
\end{equation}
\end{subequations}
\end{small}where $c_{\Delta}=\left[1,1,\cdots,-1,-1\right]$ is the difference vector, with the first half of elements as 1 and the second half as -1. 
Note that the six adaptive monopulse beams all have nulls at jamming angles and ranges, which means both MLJs and SLJs are successfully suppressed, thus they can be exploited for parameters estimation.

For elevation angle estimation, the adaptive elevation monopulse ratio is given by
\begin{small}
\begin{equation}
\hat{m}_E=\frac{\hat{f}_{\Delta_E}}{\hat{f}_{\Sigma_E}}=\frac{\left(\mathbf{w}_{\Delta_E}\right)^H\mathbf{\hat{x}}_E}{\left(\mathbf{w}_{\sum_E}\right)^H\hat{\mathbf{x}}_E}=\frac{\hat{g}_{\Delta_E}\left(\theta_s,\varphi_s\right)}{\hat{g}_{\Sigma_E}\left(\theta_s,\varphi_s\right)}
\label{Equ60}
\end{equation}
\end{small}

Since the beampattern of a rectangular array is the product of independent row and column beampatterns, the adaptive monopulse beams in elevation dimension can be expressed as
\begin{small}
\begin{subequations}
\begin{equation}
\hat{g}_{\Sigma_E}\left(\theta_s,\varphi_s\right)=g_{\Sigma_e}\left(\theta_s\right)\times\hat{g}_{\Sigma_a}\left(\varphi_s\right)
\end{equation}
\begin{equation}
\hat{g}_{\Delta_E}\left(\theta_s,\varphi_s\right)=g_{\Delta_e}\left(\theta_s\right)\times\hat{g}_{\Sigma_a}\left(\varphi_s\right)
\end{equation}
\label{Equ61}
\end{subequations}
\end{small}
The beampatterns are separated into two parts, $g$ denotes quiescent beampattern, and $\hat{g}$ denotes adaptive beampattern. Substituting (\ref{Equ61}) into (\ref{Equ60}), the adaptive elevation monopulse ratio can be further given by
\begin{small}
\begin{equation}
\hat{m}_E=\frac{\hat{g}_{\Delta_E}\left(\theta_s,\varphi_s\right)}{\hat{g}_{\Sigma_E}\left(\theta_s,\varphi_s\right)}=\frac{g_{\Delta_e}\left(\theta_s\right)}{g_{\Sigma_e}\left(\theta_s\right)}
\label{Equ62}
\end{equation}
\end{small}

The quiescent sum and difference beampatterns in the elevation dimension can be given by

\begin{small}
\begin{subequations}
\begin{equation}
g_{\Sigma_e}\left(\theta_s\right)=\left(\mathbf{w}_{\Sigma_E}\right)^Ha_{ze}\left(\theta_s\right)\notag=\frac{1-e^{j4\pi M\frac d\lambda(\sin\theta_s-\sin\theta_0)}}{1-e^{j4\pi\frac d\lambda(\sin\theta_s-\sin\theta_0)}}
\tag{63a}
\end{equation}
\begin{equation}
\begin{aligned}
g_{\Delta_e}\left(\theta_s\right)&=\left(\mathbf{w}_{\Delta_E}\right)^Ha_{ze}\left(\theta_s\right)\notag\\&=\left(1-e^{j2\pi M\frac d\lambda(\sin\theta_s-\sin\theta_0)}\right)\frac{1-e^{j2\pi M\frac d\lambda(\sin\theta_s-\sin\theta_0)}}{1-e^{j4\pi\frac d\lambda(\sin\theta_s-\sin\theta_0)}}
\end{aligned}
\tag{63b}
\end{equation}
\end{subequations}
\end{small}

Therefore, (\ref{Equ62}) can be further given by
\begin{small}
\begin{equation}
\begin{aligned}\hat{m}_{E}&=\frac{g_{\Delta_e}\left(\theta_s\right)}{g_{\Sigma_e}\left(\theta_s\right)}=\frac{1-e^{j2\pi M\frac d\lambda(\sin\theta_s-\sin\theta_0)}}{1+e^{j2\pi M\frac d\lambda(\sin\theta_s-\sin\theta_0)}}\\&\approx-j\tan\left[\pi M\frac d\lambda\cos\theta_0\left(\theta_s-\theta_0\right)\right]\end{aligned}
\label{Equ64}
\end{equation}
\end{small}

Similarly, the adaptive azimuth monopulse ratio and range monopulse ratio can be given by
\begin{small}
\begin{equation}
\hat{m}_A\approx-j\tan\left[\pi N\frac d{2\lambda}\cos\theta_0\cos\varphi_0\left(\varphi_s-\varphi_0\right)\right]
\label{Equ65}
\end{equation}
\end{small}
\begin{small}
\begin{equation}
\hat{m}_R\approx-j\tan\left[\pi\mu M\frac{\Delta t}c\big(R_s-R_0\big)\right]
\label{Equ66}
\end{equation}
\end{small}

Taking the imaginary part of these three adaptive monopulse ratios, the corresponding adaptive monopulse ratio curves can be obtained as
\begin{small}
\begin{equation}
f\left(\theta_s\right)=\mathrm{Im}\left(\hat{m}_E\right)=-\tan\left[\pi M\frac d\lambda\cos\theta_0\left(\theta_s-\theta_0\right)\right]
\label{Equ67}
\end{equation}
\begin{equation}
f\left(\varphi_s\right)=\mathrm{Im}\left(\hat{m}_A\right)=-\tan\left[\pi N\frac d{2\lambda}\mathrm{cos}\theta_0\cos\varphi_0\left(\varphi_s-\varphi_0\right)\right]
\label{Equ68}
\end{equation}
\begin{equation}
f\left(R_s\right)=\mathrm{Im}\left(\hat{m}_R\right)=-\mathrm{tan}\left[\pi\mu M\frac{\Delta t}{c}\Big(R_s-R_0\Big)\right]
\label{Equ69}
\end{equation}
\end{small}where $f\left(\theta_{s}\right), f\left(\varphi_{s}\right)$ and $f\left(R_{s}\right)$ represent the adaptive monopulse ratio curves in the elevation, azimuth and range dimensions, respectively. These three adaptive monopulse ratio curves are univariate monotonic functions near the radar array pointing direction, each solely dependent on the corresponding target parameter. Thus, the target angle and range can be estimated by the corresponding adaptive monopulse ratio values.
% Thus, we can estimate the target angle and range by calculating the corresponding adaptive monopulse ratio values.

\section{Simulation results}
\label{section:5}

To fully validate the theoretical analysis of the two proposed algorithms, simulations are provided in this section. First, we evaluate the effectiveness of MLJ suppression for the two algorithms by examining the formation of nulls in the adaptive beampatterns. Second, the results of joint angle-range estimation and root mean square error (RMSE) are provided to assess the performance of monopulse parameters estimation.

Assume a planar array consisting of $M\times N$ elements, $M=N=16$ , time shift is $\Delta{t}=1/B$ and the element spacing along the Y-axis and Z-axis is $d=0.5 \lambda$. The simulation parameters are listed in Table \ref{Table1}.
\begin{table}
\caption{Parameters of Radar\label{tab:table2}}
\centering
\begin{tabular}{llll}
			\hline
            \hline
			\textbf{Parameter} & \textbf{Value} & \textbf{Parameter}  & \textbf{Value} \\ \hline                 
            Number of rows & 16 & Number of columns & 16 \\ \hline
            Carrier frequency & 10GHz & Element spacing & 0.015m \\ \hline
            Bandwidth & 20MHz & Time shift & 0.05$\mu s$  \\ \hline  
            PRF & 20KHz & Pulse width & 10$\mu s$ \\ \hline
		\end{tabular}
        \label{Table1}
\end{table}

\subsection{Performance in jamming suppression}

The locations of target and jammings are given in Table \ref{Table2}.

\begin{table}
\caption{ Parameters of target and jammings\label{tab:table2}}
\centering
\begin{tabular}{lllll}
			\hline
            \hline
			\textbf{ } & \textbf{Targe} & \textbf{MLJ1}  & \textbf{MLJ2}  & \textbf{SLJ}\\ \hline                 
            Azimuth angle($^\circ$) & 0.5 & 3 & -3 & 8  \\ \hline
            Elevation angle($^\circ$) & 0.5 & -3 & 3 & -6 \\ \hline
            Range(km) & 75.15 & 75.3 & 74.6  & 80.8  \\ \hline  
            SNR/JNR(dB) & 10 & 40 & 40  & 40\\ \hline
		\end{tabular}
        \label{Table2}
\end{table}

(1) Experiment with four-channel adaptive beamforming algorithm

\begin{figure*}[htbp]
    \centering
    \begin{subfigure}[t]{0.24\textwidth}
        \centering
        \includegraphics[width=\textwidth]{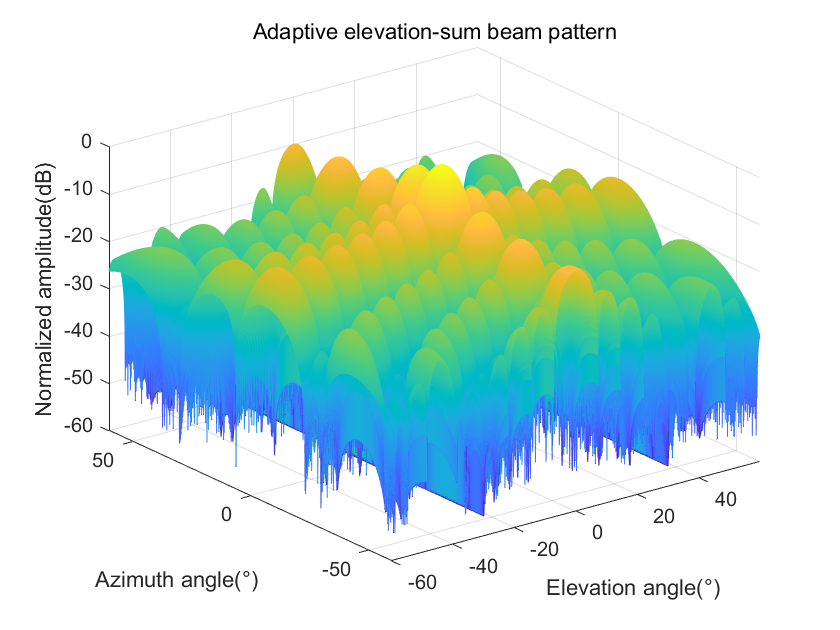}
        \caption{3-D view}
        \label{fig8:a}
    \end{subfigure}%
    \hfill
    \begin{subfigure}[t]{0.24\textwidth}
        \centering
        \includegraphics[width=\textwidth]{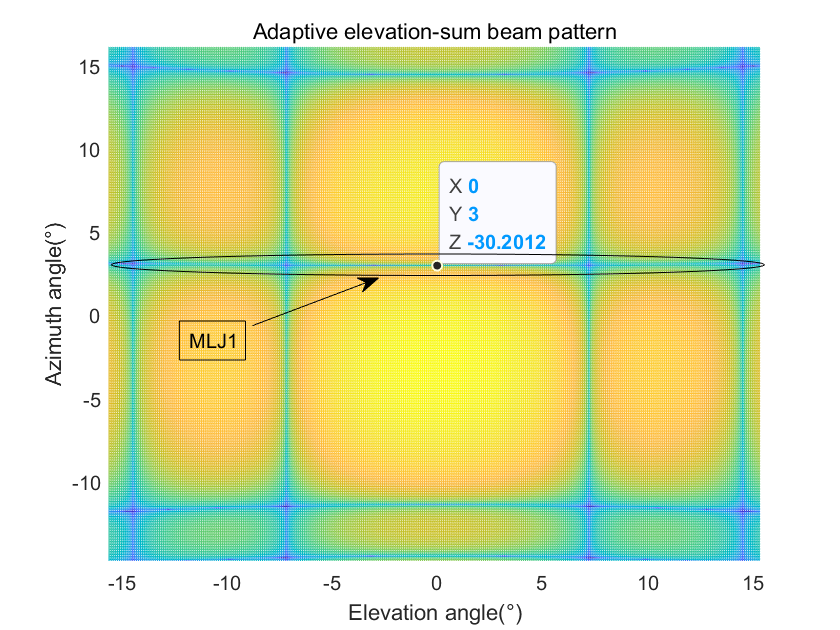}
        \caption{2-D view}
        \label{fig8:b}
    \end{subfigure} 
    \hfill
    \begin{subfigure}[t]{0.24\textwidth}
        \centering
        \includegraphics[width=\textwidth]{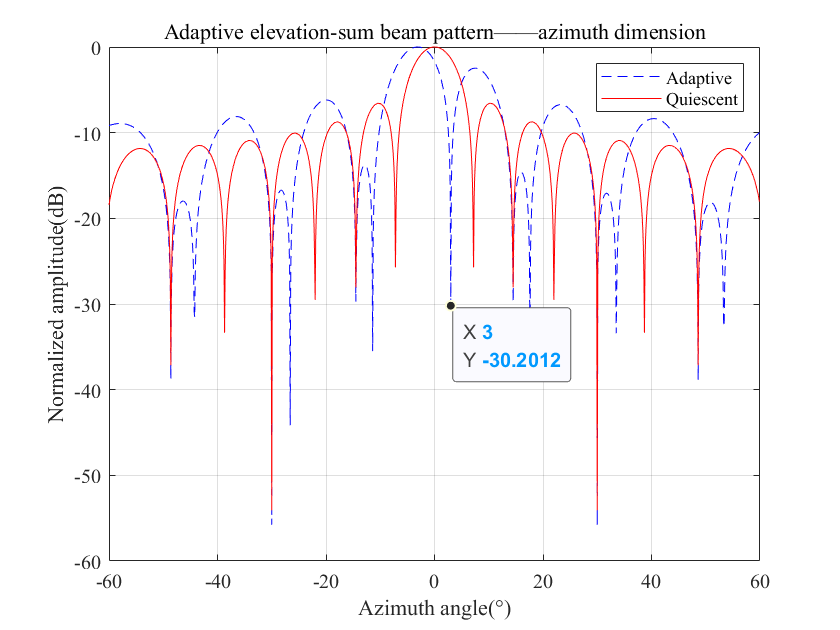}
        \caption{beampattern in azimuth dimension}
        \label{fig8:c}
    \end{subfigure}%
    \hfill
    \begin{subfigure}[t]{0.24\textwidth}
        \centering
        \includegraphics[width=\textwidth]{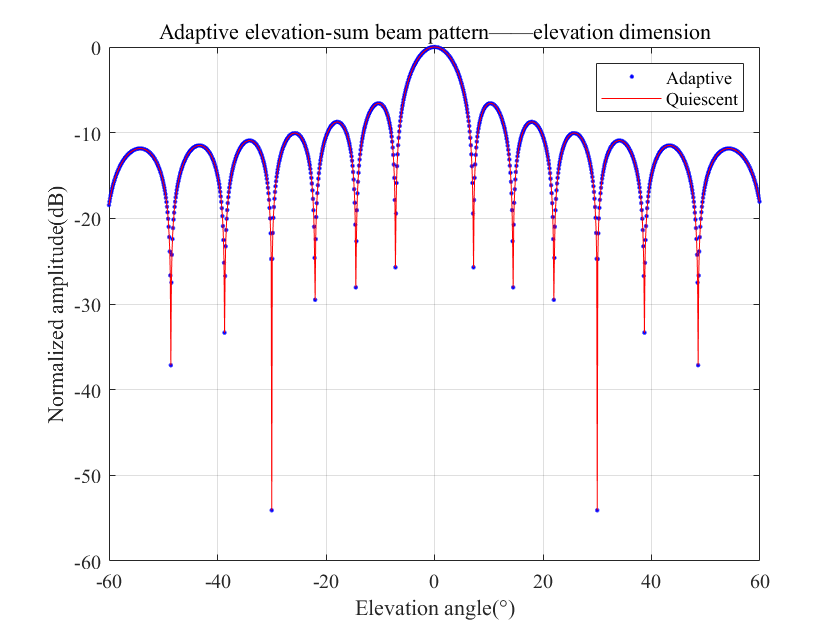}
        \caption{beampattern in elevation dimension}
        \label{fig8:d}
    \end{subfigure}

    \caption{Adaptive elevation-sum beampattern}
    \label{fig8}
\end{figure*}

In this experiment, we assume the presence of a target and MLJ1. Fig. \ref{fig8} presents the adaptive elevation-sum beampattern. Fig. \ref{fig8}(a) shows a 3-D view of the adaptive elevation-sum beampattern, while Fig. \ref{fig8}(b) depicts its 2-D view. In Fig. \ref{fig8}(b), it is evident that a null zone is formed at an azimuth angle of $3^\circ$, precisely where the MLJ1 is located, indicating effective suppression of MLJ1 along the azimuth direction. Fig. \ref{fig8}(c) compares the beampatterns in azimuth dimension of the adaptive elevation-sum beam with that of the quiescent elevation-sum beam at an elevation angle of $0^\circ$. It can be seen that a null is formed in the beampattern in azimuth dimension of the adaptive elevation-sum beam at an azimuth angle of $3^\circ$, with a depth approaching $-30dB$, providing a more intuitive demonstration that the MLJ1 has been effective suppressed along the azimuth direction. Fig. \ref{fig8}(d) compares the beampatterns in elevation dimension of the adaptive elevation-sum beam with that of the quiescent elevation-sum beam at an azimuth angle of $0^\circ $. It is evident that the beampattern in elevation dimension of the adaptive elevation-sum beam has been well maintained. To summarize, in the adaptive elevation-sum beam, the MLJ1 has been effectively suppressed along azimuth direction while keeping the beampattern undistorted along elevation direction.

\begin{figure*}[htbp]
    \centering
    \begin{subfigure}[t]{0.24\textwidth}
        \centering
        \includegraphics[width=\textwidth]{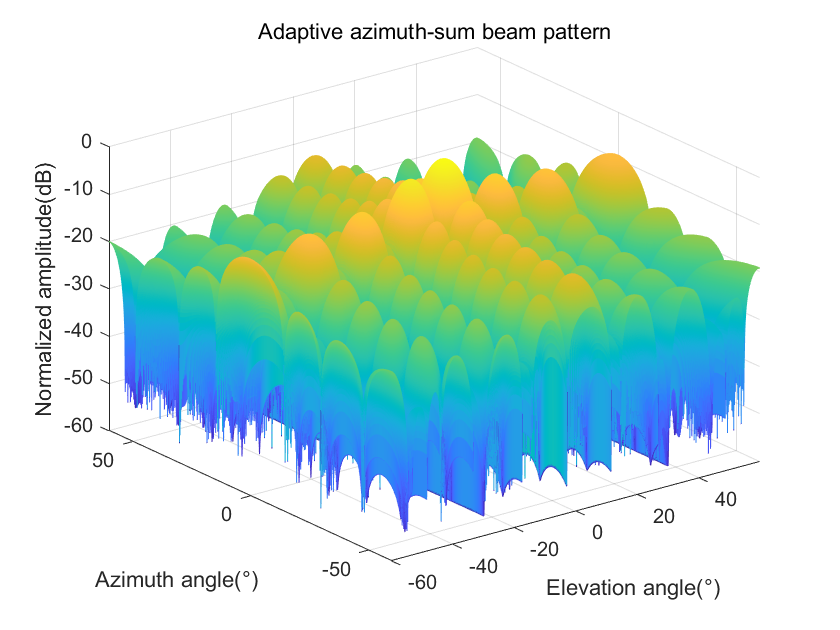}
        \caption{3-D view}
        \label{fig9:a}
    \end{subfigure}%
    \hfill
    \begin{subfigure}[t]{0.24\textwidth}
        \centering
        \includegraphics[width=\textwidth]{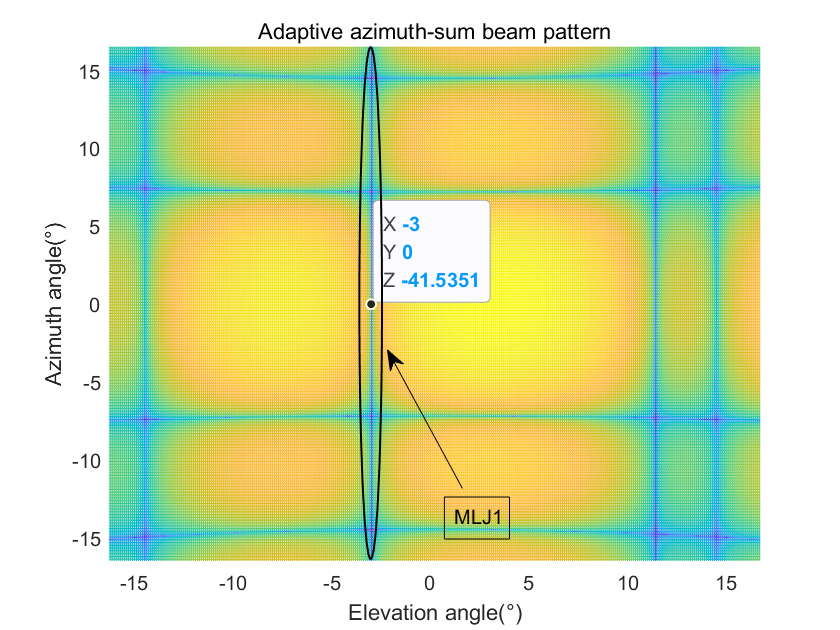}
        \caption{2-D view}
        \label{fig9:b}
    \end{subfigure} 
    \hfill
    \begin{subfigure}[t]{0.24\textwidth}
        \centering
        \includegraphics[width=\textwidth]{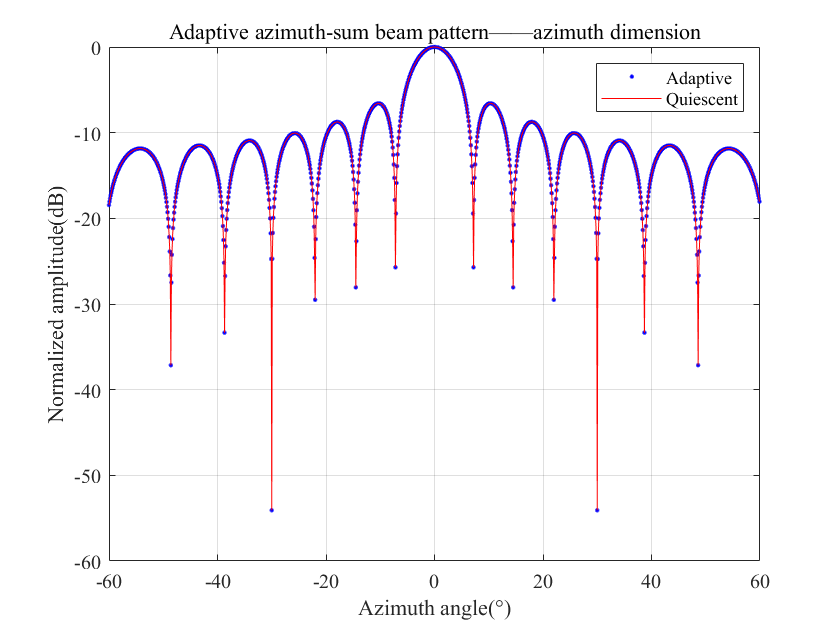}
        \caption{beampattern in azimuth dimension}
        \label{fig9:c}
    \end{subfigure}%
    \hfill
    \begin{subfigure}[t]{0.24\textwidth}
        \centering
        \includegraphics[width=\textwidth]{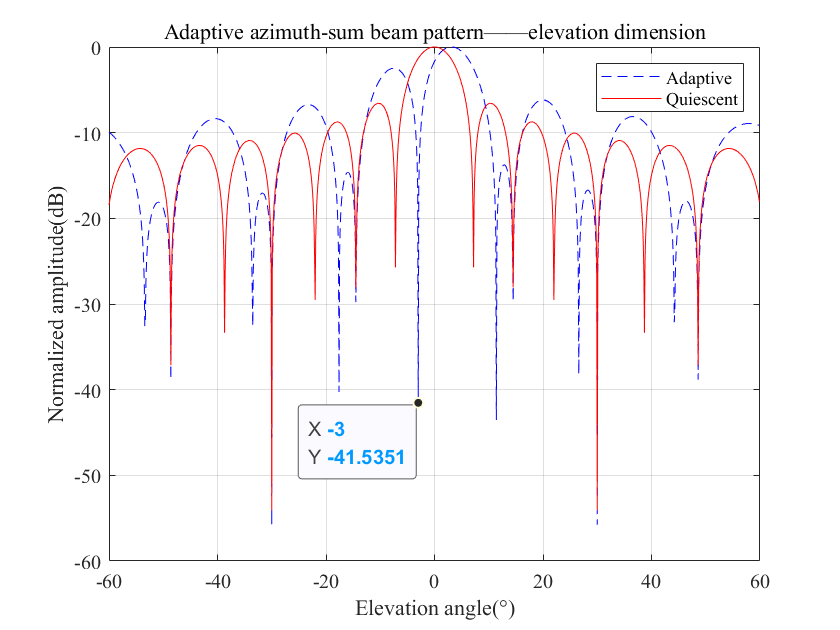}
        \caption{beampattern in elevation dimension}
        \label{fig9:d}
    \end{subfigure}

    \caption{Adaptive azimuth-sum beampattern}
    \label{fig9}
\end{figure*}

Similarly, Fig. \ref{fig9} presents the adaptive azimuth-sum beampattern. Fig. \ref{fig9}(a) shows a 3-D view of the adaptive azimuth-sum beampattern, while Fig. \ref{fig9}(b) depicts its 2-D view. In Fig. \ref{fig9}(b), it is evident that a null zone is formed at an elevation angle of $-3^\circ$, precisely where the MLJ1 is located, indicating effective suppression of MLJ1 along the elevation direction. Fig. \ref{fig9}(c) compares the beampatterns in azimuth dimension of the adaptive azimuth-sum beam with that of the quiescent azimuth-sum beam at an elevation angle of $0^\circ$. it is observed that the beampattern in azimuth dimension of the adaptive azimuth-sum beam has been well maintained. Fig. \ref{fig9}(d) compares the beampatterns in elevation dimension of the adaptive azimuth-sum beam with that of the quiescent azimuth-sum beam at an azimuth angle of $0^\circ$. It can be seen that a null is formed at an elevation angle of $-3^\circ$, with a depth approaching $-41dB$, providing a more intuitive demonstration that the MLJ1 has been effectively suppressed along the elevation direction. To summarize, in the adaptive azimuth-sum beam, the MLJ1 has been effectively canceled along elevation direction while keeping the beampattern undistorted along azimuth direction.

\begin{figure*}[htbp]
    \centering
    \begin{subfigure}[t]{0.24\textwidth}
        \centering
        \includegraphics[width=\textwidth]{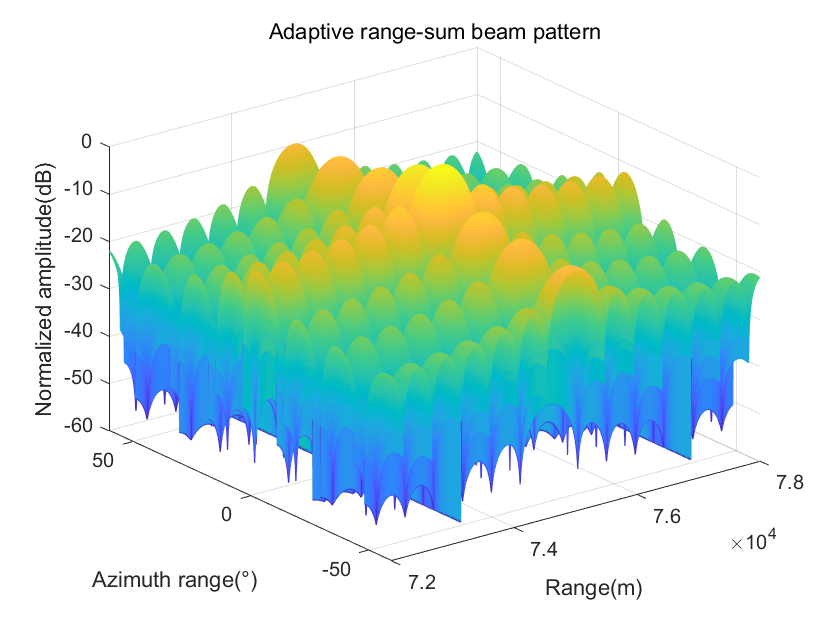}
        \caption{3-D view}
        \label{fig10:a}
    \end{subfigure}%
    \hfill
    \begin{subfigure}[t]{0.24\textwidth}
        \centering
        \includegraphics[width=\textwidth]{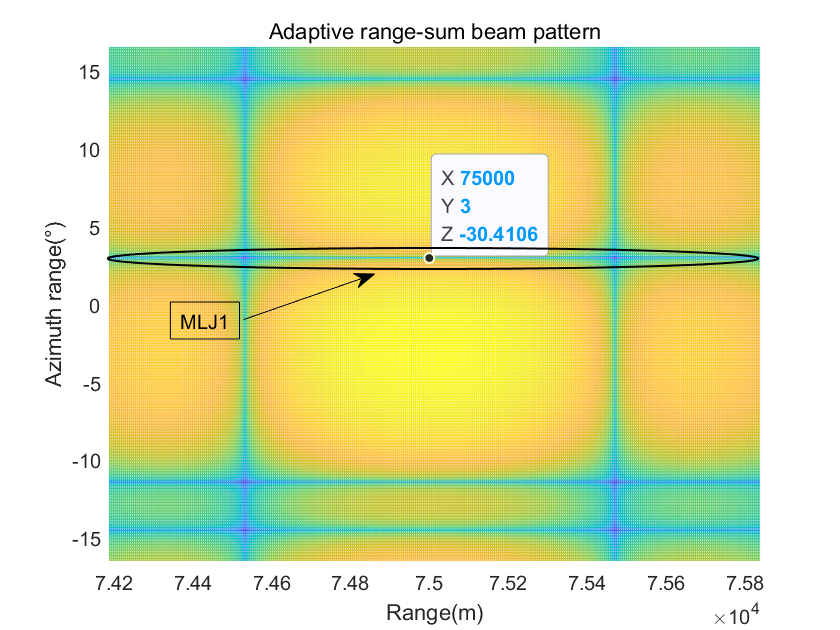}
        \caption{2-D view}
        \label{fig10:b}
    \end{subfigure} 
    \hfill
    \begin{subfigure}[t]{0.24\textwidth}
        \centering
        \includegraphics[width=\textwidth]{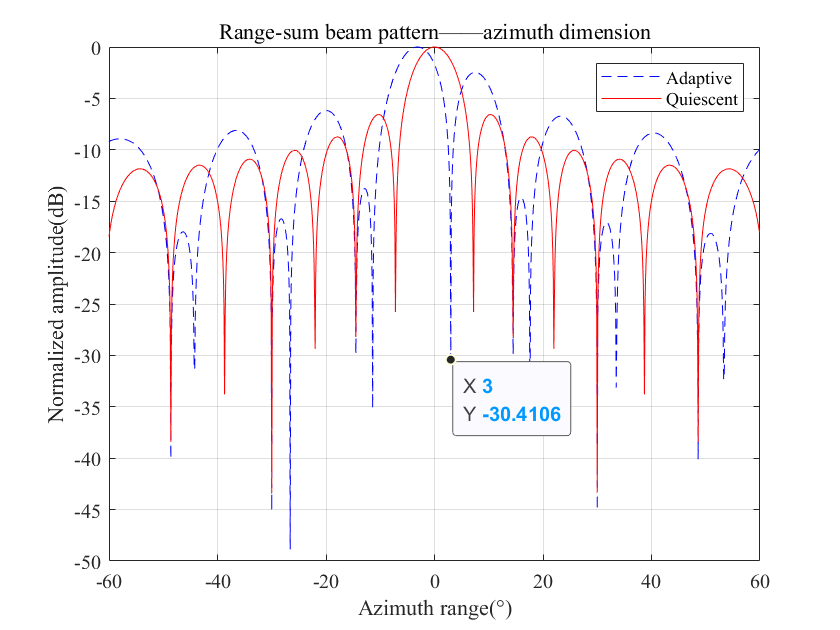}
        \caption{beampattern in azimuth dimension}
        \label{fig10:c}
    \end{subfigure}%
    \hfill
    \begin{subfigure}[t]{0.24\textwidth}
        \centering
        \includegraphics[width=\textwidth]{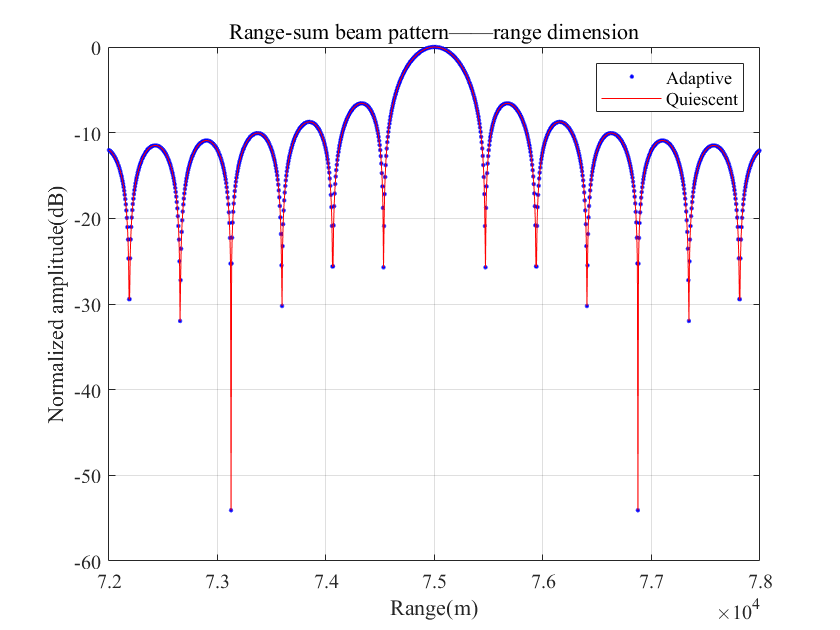}
        \caption{beampattern in range dimension}
        \label{fig10:d}
    \end{subfigure}

    \caption{Adaptive range-sum beampattern}
    \label{fig10}
\end{figure*}

After setting $\Delta{t}=1/B$, the planar array becomes an azimuth-range two-dimensional planar array. Fig. \ref{fig10} presents the adaptive range-sum beampattern. Fig. \ref{fig10}(a) shows a 3-D view of the adaptive range-sum beampattern, while Fig. \ref{fig10}(b) shows its 2-D view. In Fig. \ref{fig10}(b), it is evident that a null zone is formed at an azimuth angle of $3^\circ$, precisely where the MLJ1 is located, indicating effective suppression of MLJ1 along the azimuth direction. Fig. \ref{fig10}(c) compares the beampatterns in azimuth dimension of the adaptive range-sum beam with that of the quiescent range-sum beam at a range of $75km$. It can be seen that a null is formed in the beampattern in azimuth dimension of the adaptive range-sum beam at an azimuth angle of $3^\circ$, with a depth approaching $-30dB$, providing a more intuitive demonstration that the MLJ1 has been effectively suppressed along the azimuth direction. Fig. \ref{fig10}(d) compares the beampatterns in range dimension of the adaptive range-sum beam with that of the quiescent range-sum beam at an azimuth angle of $0^\circ$. As shown in this figure, the beampattern in range dimension of the adaptive range-sum beam has been well maintained. In brief, the MLJ1 has been effectively canceled along azimuth direction while keeping the beampattern undistorted along range direction.

(2) Experiment with row-column adaptive beamforming algorithm

In this experiment, we consider a target, two MLJs and a SLJ. The planar array is divided into $4\times4$ subarrays, with 4 elements along both the Y-axis and the Z-axis in each subarray. First, analog beamforming at element level and converting each subarray output into digital signa. Second, ABF in each row and column with four subarray beam outputs. Therefore, four DOFs have been exploited to suppress jammings, making it possible to suppress three jammings at most.

\begin{figure*}[htbp]
    \centering
    \begin{subfigure}[t]{0.24\textwidth}
        \centering
        \includegraphics[width=\textwidth]{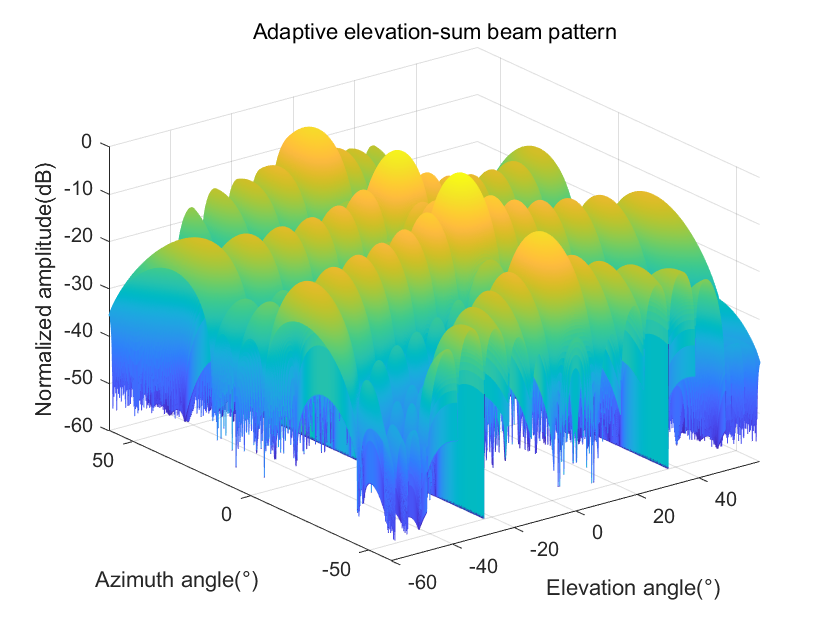}
        \caption{3-D view}
        \label{fig11:a}
    \end{subfigure}%
    \hfill
    \begin{subfigure}[t]{0.24\textwidth}
        \centering
        \includegraphics[width=\textwidth]{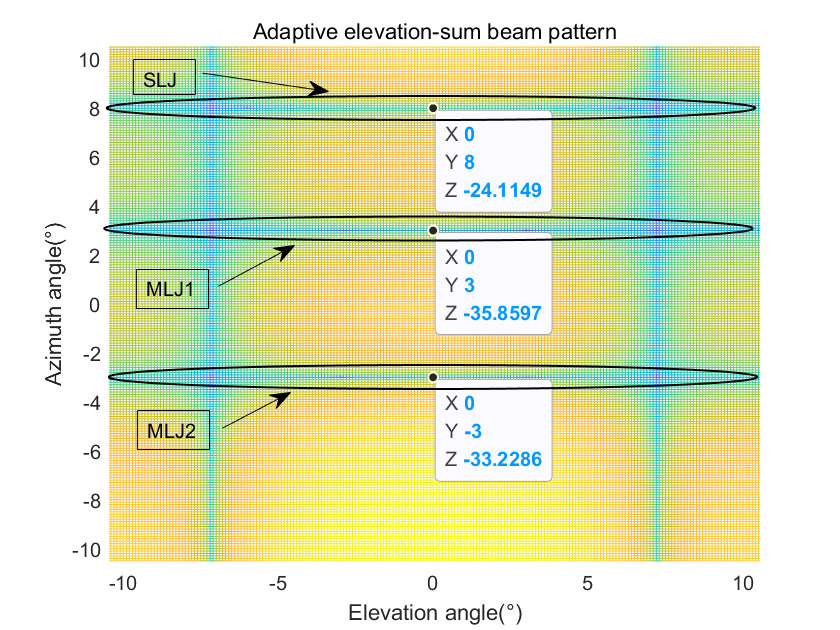}
        \caption{2-D view}
        \label{fig11:b}
    \end{subfigure} 
    \hfill
    \begin{subfigure}[t]{0.24\textwidth}
        \centering
        \includegraphics[width=\textwidth]{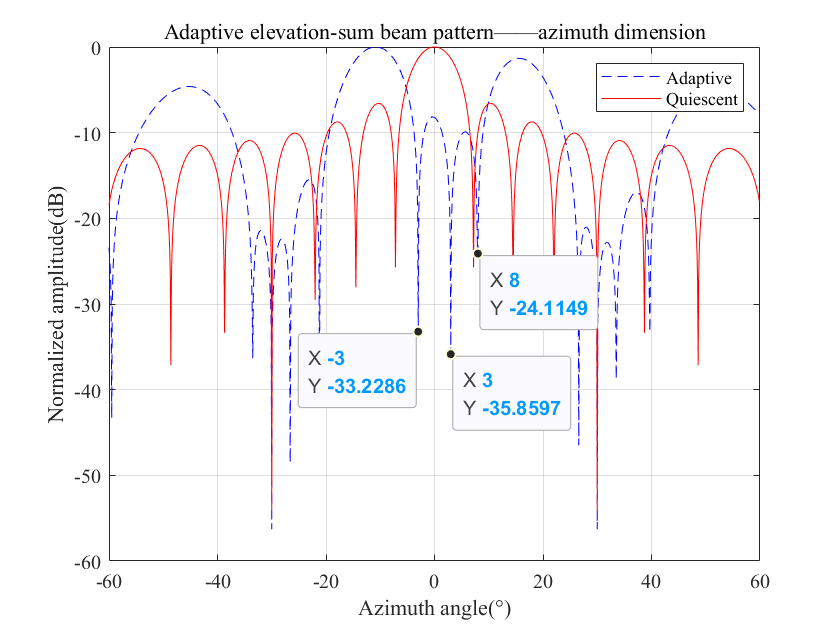}
        \caption{beampattern in azimuth dimension}
        \label{fig11:c}
    \end{subfigure}%
    \hfill
    \begin{subfigure}[t]{0.24\textwidth}
        \centering
        \includegraphics[width=\textwidth]{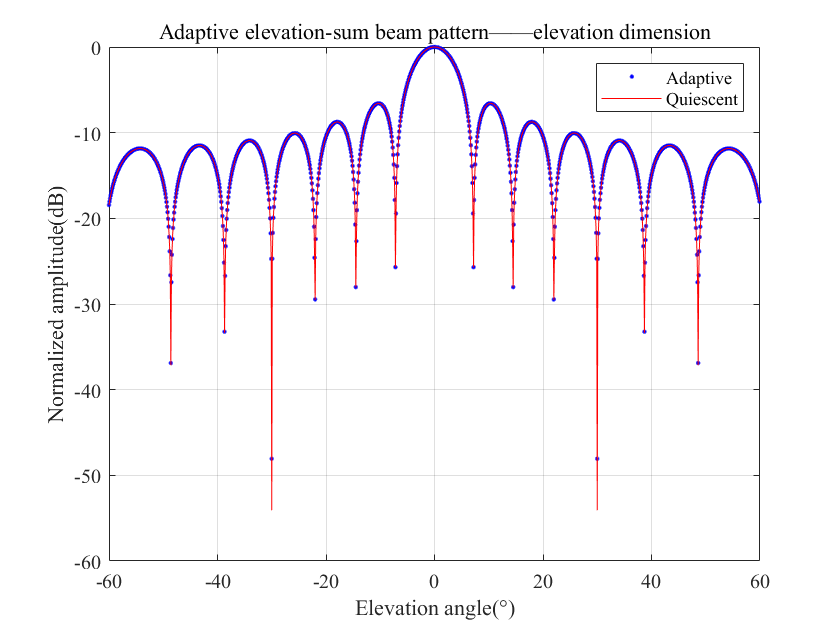}
        \caption{beampattern in elevation dimension}
        \label{fig11:d}
    \end{subfigure}

    \caption{Adaptive elevation-sum beampattern}
    \label{fig11}
\end{figure*}

Fig. \ref{fig11} presents the adaptive elevation-sum beampattern. Fig. \ref{fig11}(a) shows a 3-D view of the adaptive elevation-sum beampattern, while Fig. \ref{fig11}(b) shows its 2-D view. In Fig. \ref{fig11}(b), it is evident that three null zones are formed at azimuth angles of $-3^\circ$, $3^\circ$ and $8^\circ$, precisely where the two MLJs and a SLJ are located. This indicates effective suppression of multiple jammings along the azimuth direction. Fig. \ref{fig11}(c) compares the beampatterns in azimuth dimension of the adaptive elevation-sum beam with that of the quiescent elevation-sum beam at an elevation angle of $0^\circ$. It can be seen that three nulls are formed in the beampattern in azimuth dimension of the adaptive elevation-sum beam at azimuth angles of $-3^\circ$, $3^\circ$ and $8^\circ$, providing a more intuitive demonstration that the multiple jammings have been effectively suppressed along the azimuth direction. Fig. \ref{fig11}(d) compares the beampatterns in elevation dimension of the adaptive elevation-sum beam with that of the quiescent elevation-sum beam at an azimuth angle of $0^\circ$. It is evident that the beampattern in elevation dimension of the adaptive elevation-sum beam has been well maintained. To sum up, in the adaptive elevation-sum beam, the multiple jammings have been effectively canceled along azimuth direction while keeping the beampattern undistorted along elevation direction.

\begin{figure*}[htbp]
    \centering
    \begin{subfigure}[t]{0.24\textwidth}
        \centering
        \includegraphics[width=\textwidth]{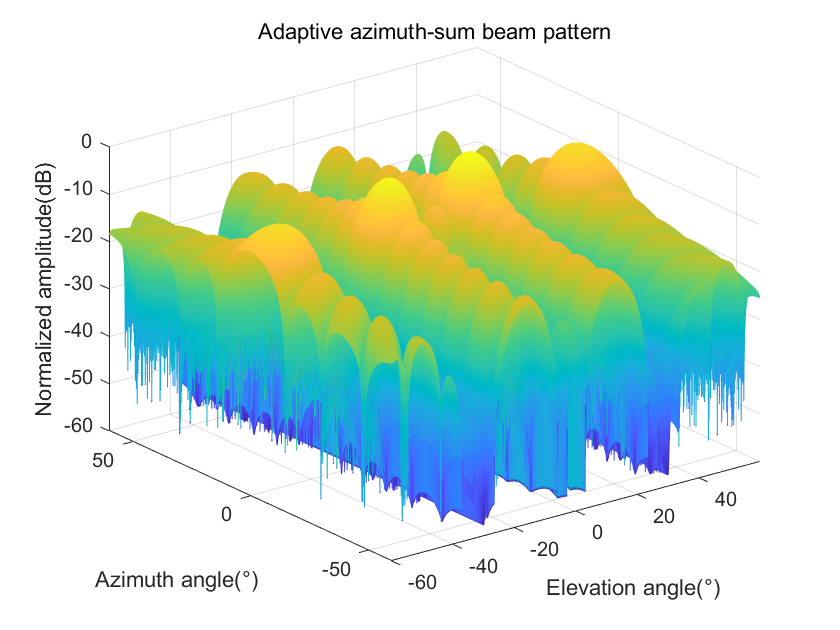}
        \caption{3-D view}
        \label{fig12:a}
    \end{subfigure}%
    \hfill
    \begin{subfigure}[t]{0.24\textwidth}
        \centering
        \includegraphics[width=\textwidth]{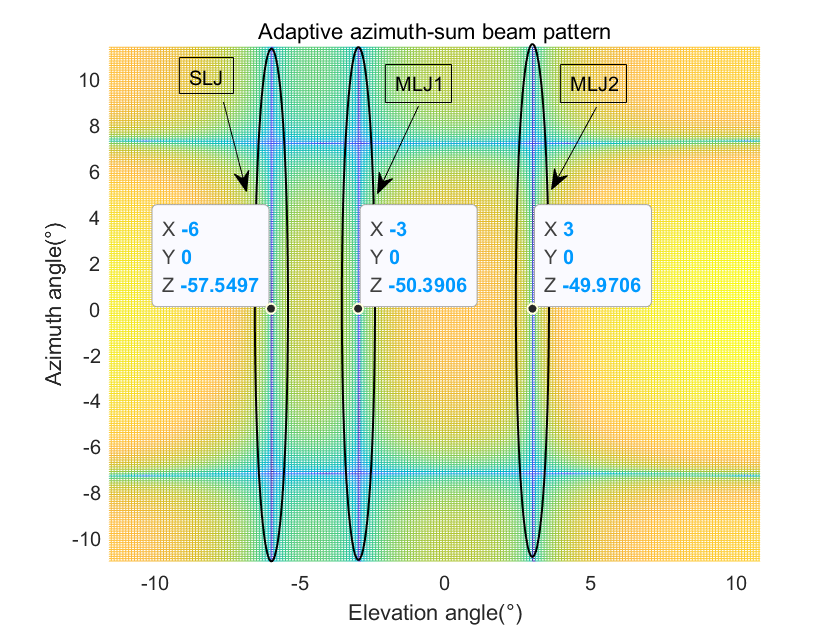}
        \caption{2-D view}
        \label{fig12:b}
    \end{subfigure} 
    \hfill
    \begin{subfigure}[t]{0.24\textwidth}
        \centering
        \includegraphics[width=\textwidth]{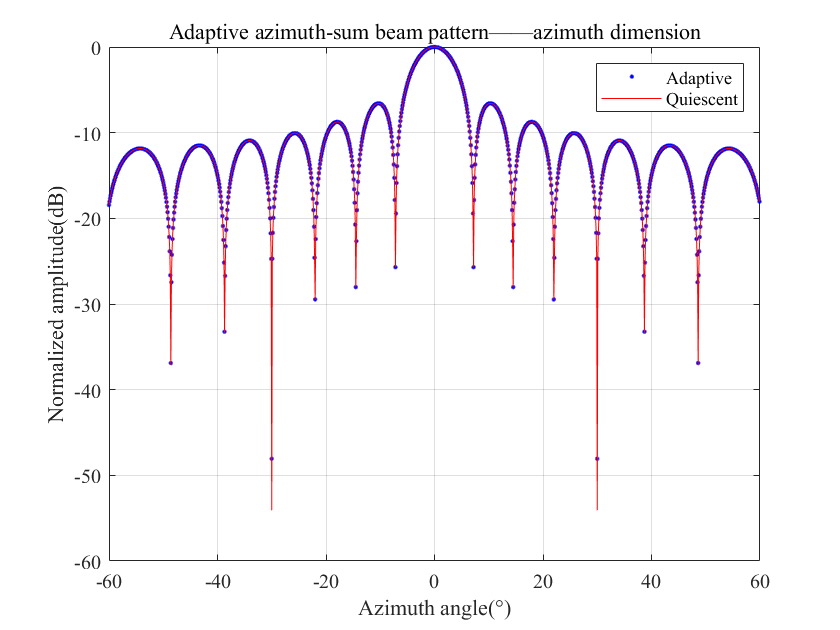}
        \caption{beampattern in azimuth dimension}
        \label{fig12:c}
    \end{subfigure}%
    \hfill
    \begin{subfigure}[t]{0.24\textwidth}
        \centering
        \includegraphics[width=\textwidth]{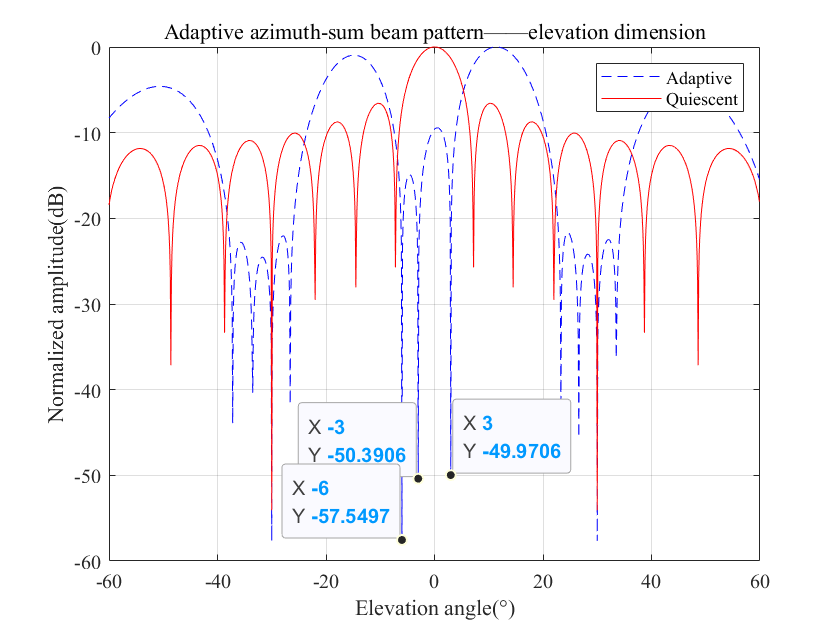}
        \caption{beampattern in elevation dimension}
        \label{fig12:d}
    \end{subfigure}

    \caption{Adaptive azimuth-sum beampattern}
    \label{fig12}
\end{figure*}

Fig. \ref{fig12} presents the adaptive azimuth-sum beampattern. Fig. \ref{fig12}(a) shows a 3-D view of the adaptive azimuth-sum beampattern, while Fig. \ref{fig12}(b) depicts its 2-D view. In Fig. \ref{fig12}(b), it is evident that three null zones are formed at elevation angles of $-6^\circ$, $-3^\circ$and $3^\circ$, precisely where the two MLJs and a SLJ are located, indicating effective suppression of multiple jammings along the elevation direction. Fig. \ref{fig12}(c) compares the beampatterns in azimuth dimension of the adaptive azimuth-sum beam with that of the quiescent azimuth-sum beam at an elevation angle of $0^\circ$. As shown, the beampattern in azimuth dimension of the adaptive azimuth-sum beam has been well maintained. Fig. \ref{fig12}(d) compares the beampatterns in elevation dimension of the adaptive azimuth-sum beam with that of the quiescent azimuth-sum beam at an azimuth angle of $0^\circ$. It can be seen that three nulls are formed in the beampattern in elevation dimension of the adaptive azimuth-sum beam at elevation angles of $-6^\circ$, $-3^\circ$ and $3^\circ$. This further demonstrates that the multiple jammings have been effectively suppressed along the elevation direction. To summarize, in the adaptive azimuth-sum beam, the multiple jammings have been effectively canceled along elevation direction while keeping the beampattern undistorted along azimuth direction.

\begin{figure*}[htbp]
    \centering
    \begin{subfigure}[t]{0.24\textwidth}
        \centering
        \includegraphics[width=\textwidth]{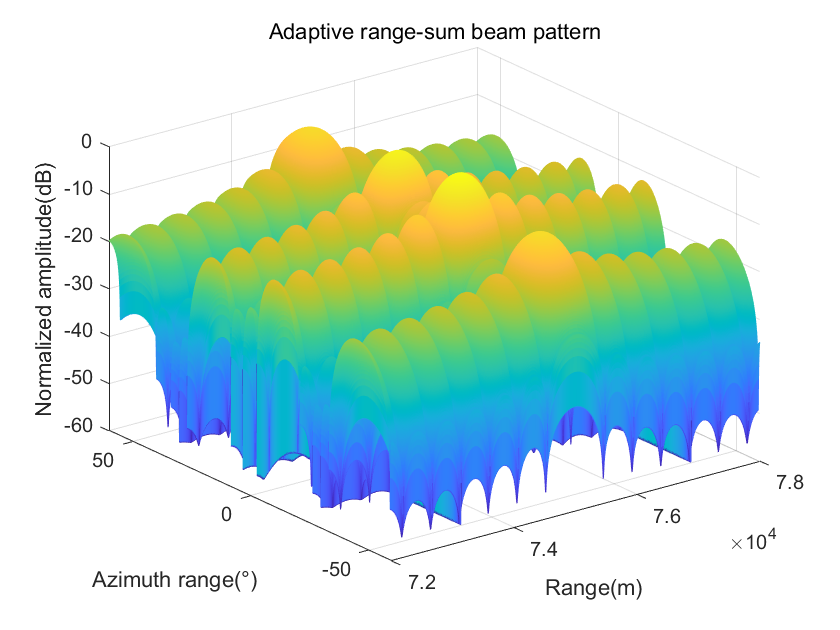}
        \caption{3-D view}
        \label{fig13:a}
    \end{subfigure}%
    \hfill
    \begin{subfigure}[t]{0.24\textwidth}
        \centering
        \includegraphics[width=\textwidth]{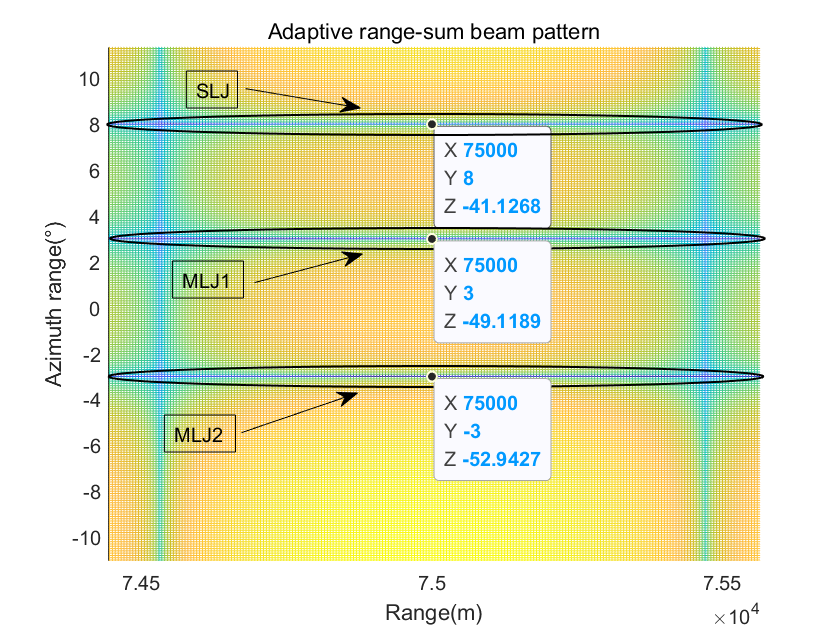}
        \caption{2-D view}
        \label{fig13:b}
    \end{subfigure} 
    \hfill
    \begin{subfigure}[t]{0.24\textwidth}
        \centering
        \includegraphics[width=\textwidth]{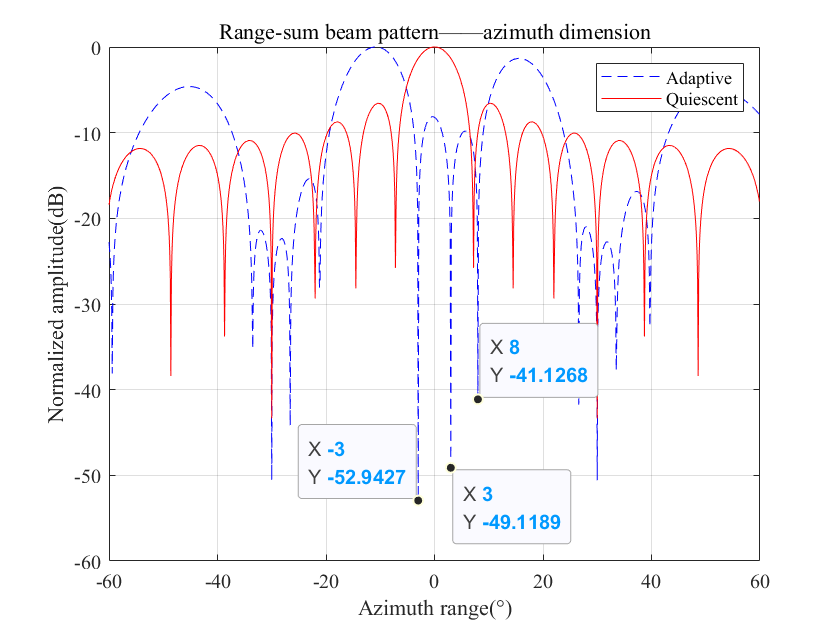}
        \caption{beampattern in azimuth dimension}
        \label{fig13:c}
    \end{subfigure}%
    \hfill
    \begin{subfigure}[t]{0.24\textwidth}
        \centering
        \includegraphics[width=\textwidth]{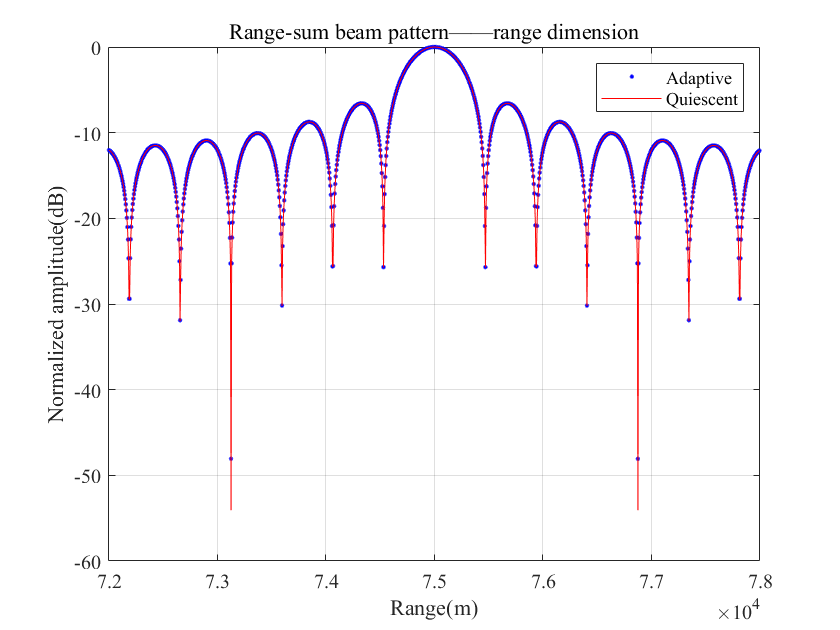}
        \caption{beampattern in range dimension}
        \label{fig13:d}
    \end{subfigure}

    \caption{Adaptive range-sum beampattern}
    \label{fig13}
\end{figure*}

After setting $\Delta{t}=1/B$, the planar array becomes an azimuth-range two-dimensional planar array. Fig.\ref{fig13} presents the adaptive range-sum beampattern. Fig. \ref{fig13}(a) shows a 3-D view of the adaptive range-sum beampattern, while Fig. \ref{fig13}(b) depicts its 2-D view. In Fig. \ref{fig13}(b), three null zones are formed at azimuth angles of $-3^\circ$, $3^\circ$ and $8^\circ$, precisely where the two MLJs and a SLJ are located, indicating effective suppression of multiple jammings along the azimuth direction. Fig. \ref{fig13}(c) compares the beampatterns in azimuth dimension of the adaptive range-sum beam with that of the quiescent range-sum beam at a range of $75km$. It can be seen that three nulls are formed in the beampattern in azimuth dimension of the adaptive range-sum beam at azimuth angles of $-3^\circ$, $3^\circ$ and $8^\circ$. This further clarifies that the multiple jammings have been effectively suppressed along the azimuth direction. Fig. \ref{fig13}(d) compares the beampatterns in range dimension of the adaptive range-sum beam with that of the quiescent range-sum beam at an azimuth angle of $0^\circ$. It is evident that the beampattern in range dimension of the adaptive range-sum beam has been well maintained. In conclusion, the multiple jammings have been effectively suppressed along azimuth direction while keeping the beampattern undistorted along range direction.

\subsection{Performance in parameters estimation}

After suppressing jamming, the adaptive sum-and-difference beams are used for monopulse parameters estimation. The target angle or range is then retrieved by mapping the adaptive monopulse ratio to the corresponding curve.
% After calculating adaptive monopulse ratio value, the angle or range parameter can be indexed from corresponding monopulse ratio curve.

\begin{figure*}[htbp]
    \centering
    \begin{subfigure}[t]{0.3\textwidth}
        \centering
        \includegraphics[width=\textwidth]{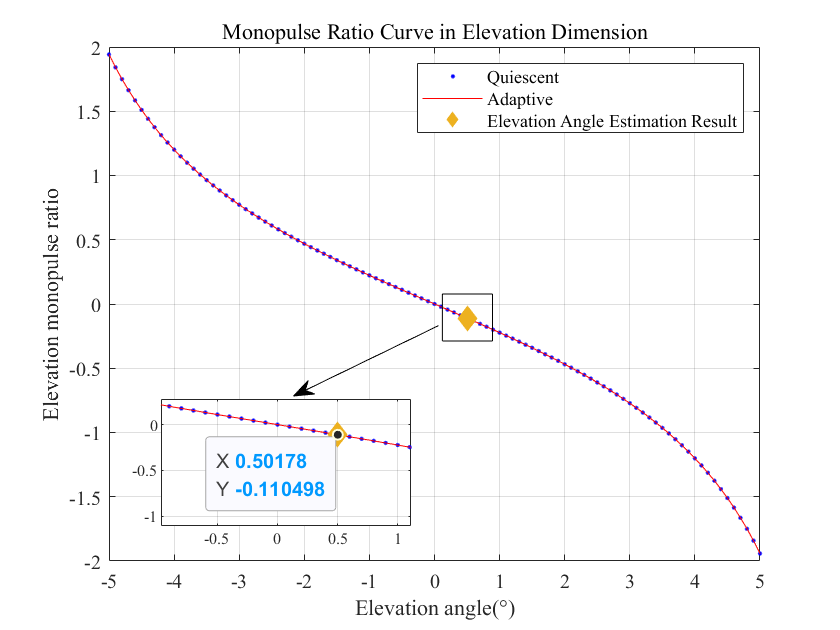}
        \caption{elevation angle}
        \label{fig14:a}
    \end{subfigure}%
    \hfill
    \begin{subfigure}[t]{0.3\textwidth}
        \centering
        \includegraphics[width=\textwidth]{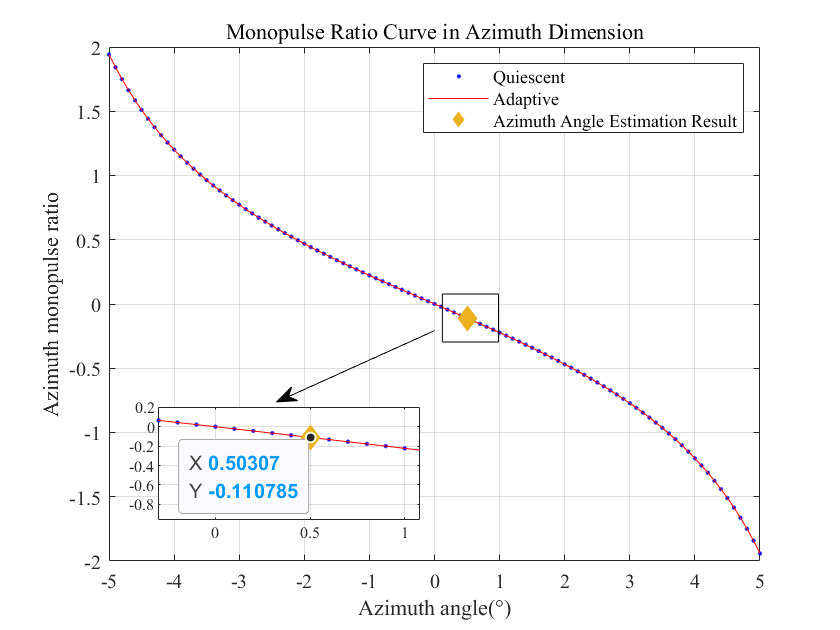}
        \caption{azimuth angle}
        \label{fig14:b}
    \end{subfigure} 
    \hfill
    \begin{subfigure}[t]{0.3\textwidth}
        \centering
        \includegraphics[width=\textwidth]{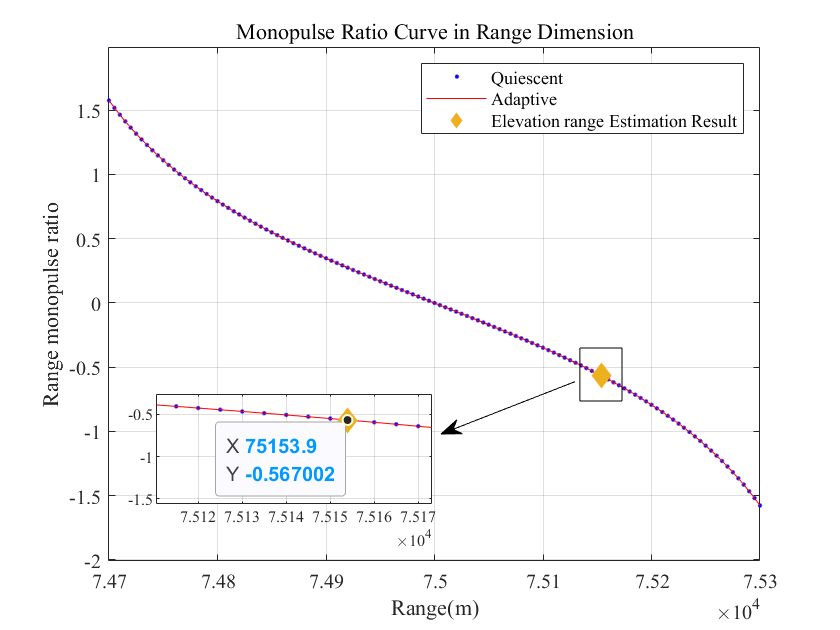}
        \caption{range}
        \label{fig14:c}
    \end{subfigure}%

    \caption{Monopulse ratio curve and parameters estimation result using four-channel ABF algorithm}
    \label{fig14}
\end{figure*}

Fig. \ref{fig14} compares the quiescent monopulse ratio curves with the adaptive monopulse ratio curves obtained using the four-channel ABF algorithm in the elevation, azimuth and range dimensions, respectively, along with the corresponding parameter estimation results. In Fig. \ref{fig14}(a), the quiescent and adaptive monopulse ratio curves in elevation dimension are completely consistent, indicating that MLJ suppression does not cause distortion in monopulse ratio. Additionally, the elevation angle estimation result is $0.50178^\circ$, with an error of $0.00178^\circ$ compared to the true target elevation angle. Similarly, in Fig. \ref{fig14}(b) and Fig. \ref{fig14}(c), the adaptive azimuth angle monopulse ratio curve is maintained, as is the adaptive range monopulse ratio curve. The azimuth angle estimation result is $0.50307^\circ$, the range estimation result is $75.1539km$, both very close to the parameters of the target location.

\begin{figure*}[htbp]
    \centering
    \begin{subfigure}[t]{0.3\textwidth}
        \centering
        \includegraphics[width=\textwidth]{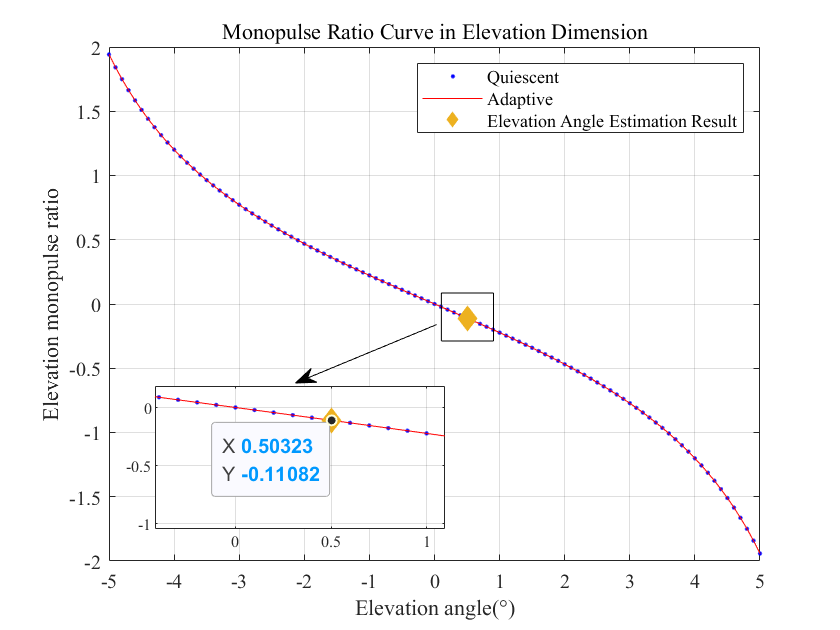}
        \caption{elevation angle}
        \label{fig15:a}
    \end{subfigure}%
    \hfill
    \begin{subfigure}[t]{0.3\textwidth}
        \centering
        \includegraphics[width=\textwidth]{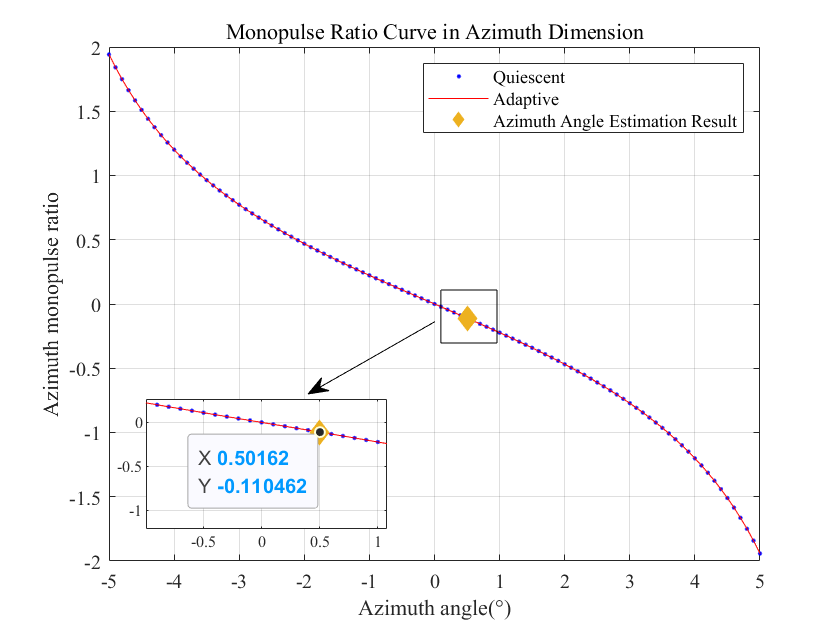}
        \caption{azimuth angle}
        \label{fig15:b}
    \end{subfigure} 
    \hfill
    \begin{subfigure}[t]{0.3\textwidth}
        \centering
        \includegraphics[width=\textwidth]{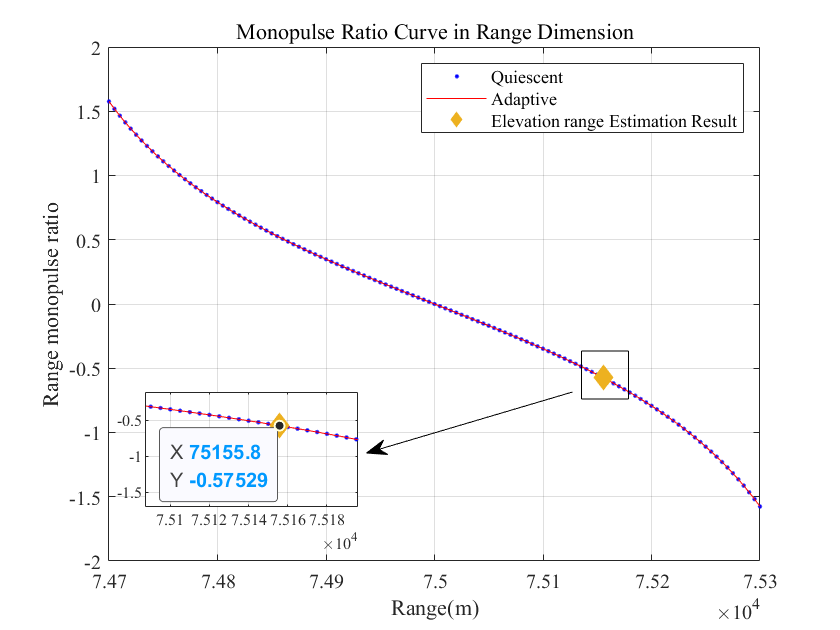}
        \caption{range}
        \label{fig15:c}
    \end{subfigure}%

    \caption{Monopulse ratio curve and parameters estimation result using row-column ABF algorithm}
    \label{fig15}
\end{figure*}

\begin{figure*}[htbp]
    \centering
    \begin{subfigure}[t]{0.3\textwidth}
        \centering
        \includegraphics[width=\textwidth]{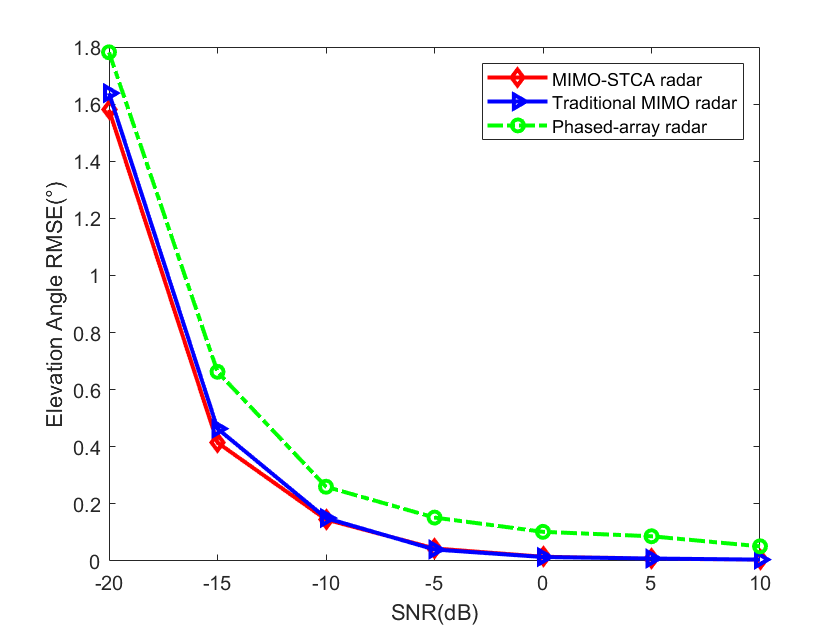}
        \caption{elevation angle}
        \label{fig16:a}
    \end{subfigure}%
    \hfill
    \begin{subfigure}[t]{0.3\textwidth}
        \centering
        \includegraphics[width=\textwidth]{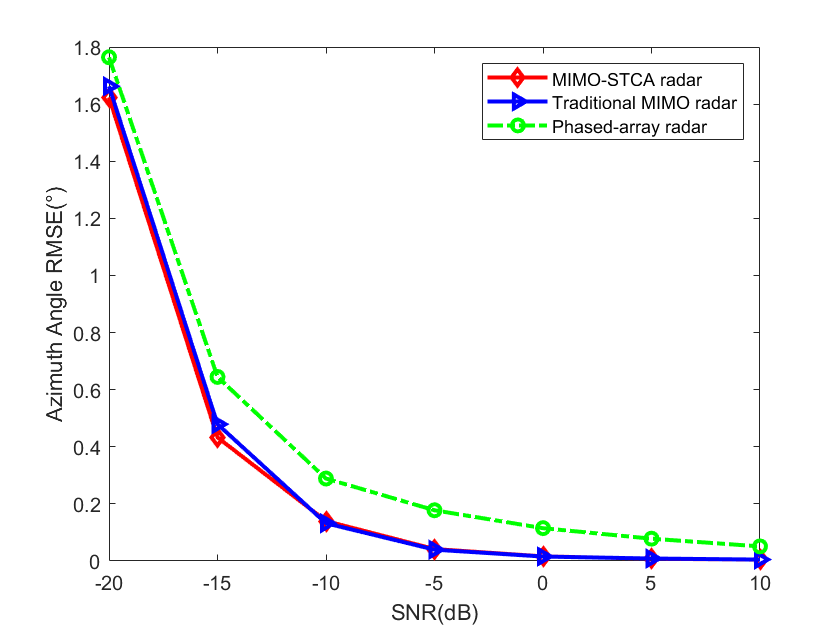}
        \caption{azimuth angle}
        \label{fig16:b}
    \end{subfigure} 
    \hfill
    \begin{subfigure}[t]{0.3\textwidth}
        \centering
        \includegraphics[width=\textwidth]{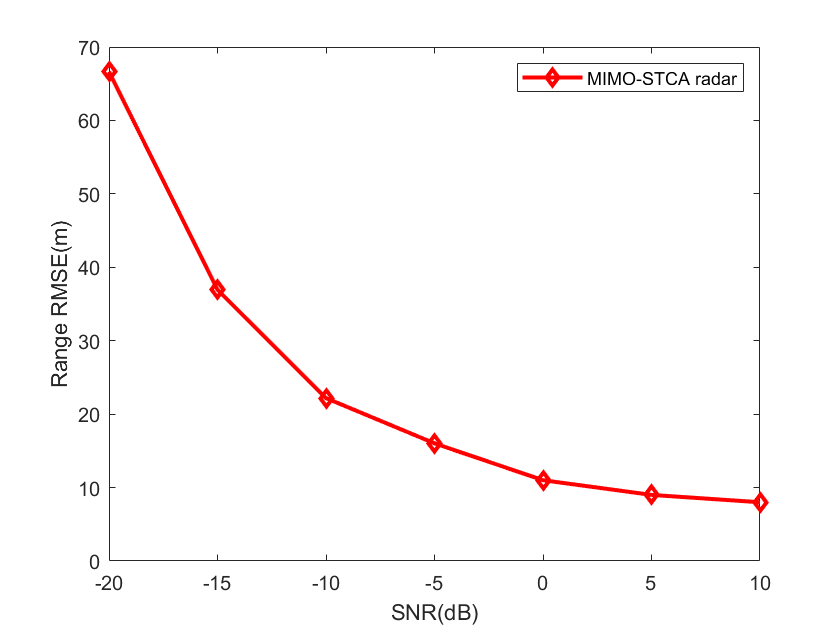}
        \caption{range}
        \label{fig16:c}
    \end{subfigure}%

    \caption{RMSEs of parameter estimation versus SNR}
    \label{fig16}
\end{figure*}

Simlar to Fig. \ref{fig14}, Fig. \ref{fig15} compares the quiescent monopulse ratio curves with the adaptive monopulse ratio curves obtained using the row-column ABF algorithm. It can be seen that the quiescent and adaptive monopulse ratio curves in all three dimensions are completely consistent, indicating that multiple jammings suppression does not cause distortion in monopulse ratio. Further, the estimation result for the elevation angle, azimuth angle and range are $0.50323^\circ$, $0.50162^\circ$ and $75.1558km$, respectively, with errors of $0.00323^\circ$, $0.00162^\circ$ and $5.8m$ compared to the true parameters of the target.

% Moreover, to further validate the accuracy of parameter estimation and analyze the influence of different SNRs on the performance of parameter estimation results, the RMSEs of angle estimation and range estimation are showm in Fig. \ref{fig16}, which can be defined as
Moreover, the RMSEs of angle estimation and range estimation are showm in Fig. \ref{fig16}, which can be defined as
\begin{small}
\begin{subequations}
\begin{equation}
RMSE_\theta=\sqrt{\frac1L\sum_{l=1}^L\left(\hat{\theta}_l-\theta_s\right)^2}
\end{equation}
\begin{equation}
RMSE_\varphi=\sqrt{\frac{1}{L}\sum_{l=1}^L\left(\hat{\varphi}_l-\varphi_s\right)^2}
\end{equation}
\begin{equation}
RMSE_R=\sqrt{\frac{1}{L}\sum_{l=1}^L\left(\hat{R}_l-R_s\right)^2}
\end{equation}
\end{subequations}
\end{small}where $\hat{\theta}_{l},\hat{\varphi}_{l}$ and $\hat{R}_{l}$ denote the estimated elevation angle, azimuth angle and range in the $lth$ Monte-Carlo experiment, and $L$ denotes the total number of Monte-Carlo experiments. This study conducts 1000 Monte-Carlo experiments.
% In this experiment, 1000 Monte-Carlo experiments are performed.

As shown in Fig. \ref{fig16}(a) and Fig. \ref{fig16}(b), it can be clearly seen that the RMSE of angle estimation decreases as the SNR increases. The RMSEs of the traditonal MIMO radar and the MIOM-STCA radar are similar and both outperform that of the Phased-array radar. Since the traditional MIMO radar and the Phased-array radar cannot extract range information, monopulse technique cannot be applied for range parameter estimation. Therefore, only the RMSE of range estimation with MIMO-STCA radar is shown in Fig. \ref{fig16}(c). It is evident that the RMSE of range estimation also decreases with the increase in SNR, and stays at a relatively low error level.

\begin{figure}
    \centering  %图片全局居中
    \subfloat[]{
    \label{Fig.sub.1}
    \includegraphics[width=0.5\columnwidth]{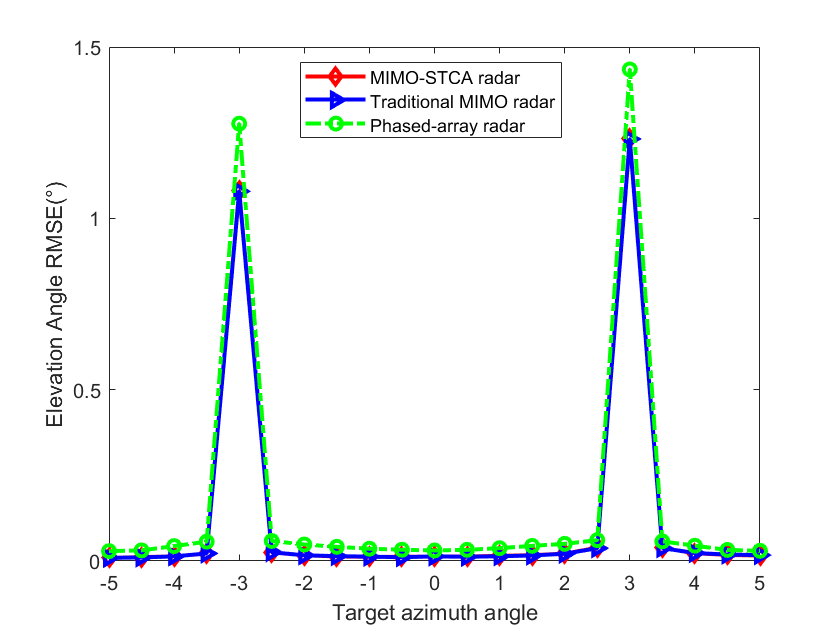}}
    \subfloat[]{
    \label{Fig.sub.2}
    \includegraphics[width=0.5\columnwidth]{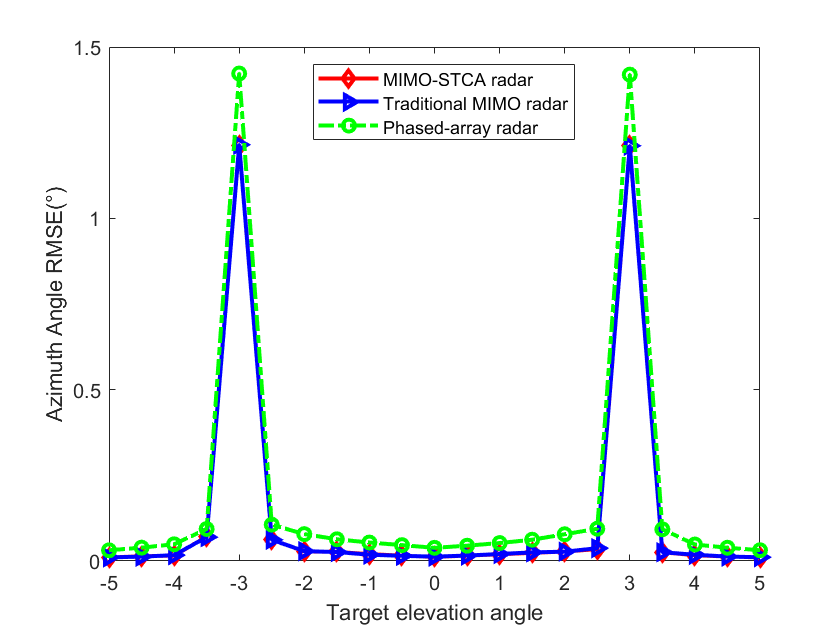}}
    
    \caption{RMSEs of parameter estimation vary with the target position (a) elevation angle (b) azimuth angle}
    \label{fig17}
\end{figure}

Fig. \ref{fig17} shows the results of the angle estimation RMSE as the target angle changes, in the presence of MLJ1 and MLJ2. It can be observed that the angle estimation error increases as the target approaches the MLJ. Notably, the performance of the traditional MIMO radar and the MIMO-STCA radar are similar, both outperforming the Phased-array radar.

% \begin{figure*}[htbp]
%     \centering
%     \begin{subfigure}[t]{0.25\textwidth}
%         \centering
%         \includegraphics[width=\textwidth]{RMSE of elevation angle estimation.png}
%         \caption{elevation angle}
%         \label{fig17:a}
%     \end{subfigure}%
%     \hfill
%     \begin{subfigure}[t]{0.25\textwidth}
%         \centering
%         \includegraphics[width=\textwidth]{RMSE of azimuth angle estimation.png}
%         \caption{azimuth angle}
%         \label{fig17:b}
%     \end{subfigure} 

%     \caption{RMSEs of parameter estimation vary with the target position}
%     \label{fig17}
% \end{figure*}

\section{Conclusions}
\label{section:6}

% \begin{figure}[htbp]  % 普通 figure 自动适应单栏
%     \centering
%     % 第一张子图
%     \begin{subfigure}[b]{\linewidth}  % 让子图占满单栏
%         \centering
%         \includegraphics[width=0.9\linewidth]{RMSE of elevation angle estimation.png}
%         \caption{elevation angle}
%         \label{fig17:a}
%     \end{subfigure}
%     % 间距
%     \vspace{5pt}
%     % 第二张子图
%     \begin{subfigure}[b]{\linewidth}
%         \centering
%         \includegraphics[width=0.9\linewidth]{RMSE of azimuth angle estimation.png}
%         \caption{azimuth angle}
%         \label{fig17:b}
%     \end{subfigure}
    
%     % 总体标题
%     \caption{RMSEs of parameter estimation vary with the target position}
%     \label{fig17}
% \end{figure}

In this paper, we addressed the challenges of mainlobe jamming  suppression and joint angle-range estimation in monopulse radar systems by proposing a novel MIMO-STCA framework integrated with adaptive beamforming techniques. To achieve accurate parameter estimation under complex jamming scenarios, we developed two key algorithms: a four-channel ABF method that introduces an additional delta-delta channel for single MLJ suppression while preserving monopulse ratio integrity, and a hierarchical row-column ABF algorithm that leverages subarray-level spatial DOFs to simultaneously mitigate multiple MLJs and SLJs. The proposed MIMO-STCA architecture exploits time-shifted transmissions across array rows to create range-domain DOFs, enabling joint angle-range estimation through monopulse processing. Theoretical analysis confirmed that both algorithms maintain undistorted sum and difference beampatterns along orthogonal spatial directions during interference suppression, while numerical simulations demonstrated precise null steering toward jammer locations and consistent monopulse ratio preservation. Experimental validation showed significant improvements in estimation accuracy, with RMSEs reduced by over 10\%-60\%  compared to the Phased-array radar in MLJ scenarios. Future research directions may focus on real-time implementation of the proposed algorithms for dynamic jamming environments and extension of the framework to multi-target scenarios.

\bibliographystyle{IEEEtran}
\bibliography{reference}

\vspace{-60pt}
\begin{IEEEbiography}[{\includegraphics[width=1in,height=1.25in,clip,keepaspectratio]{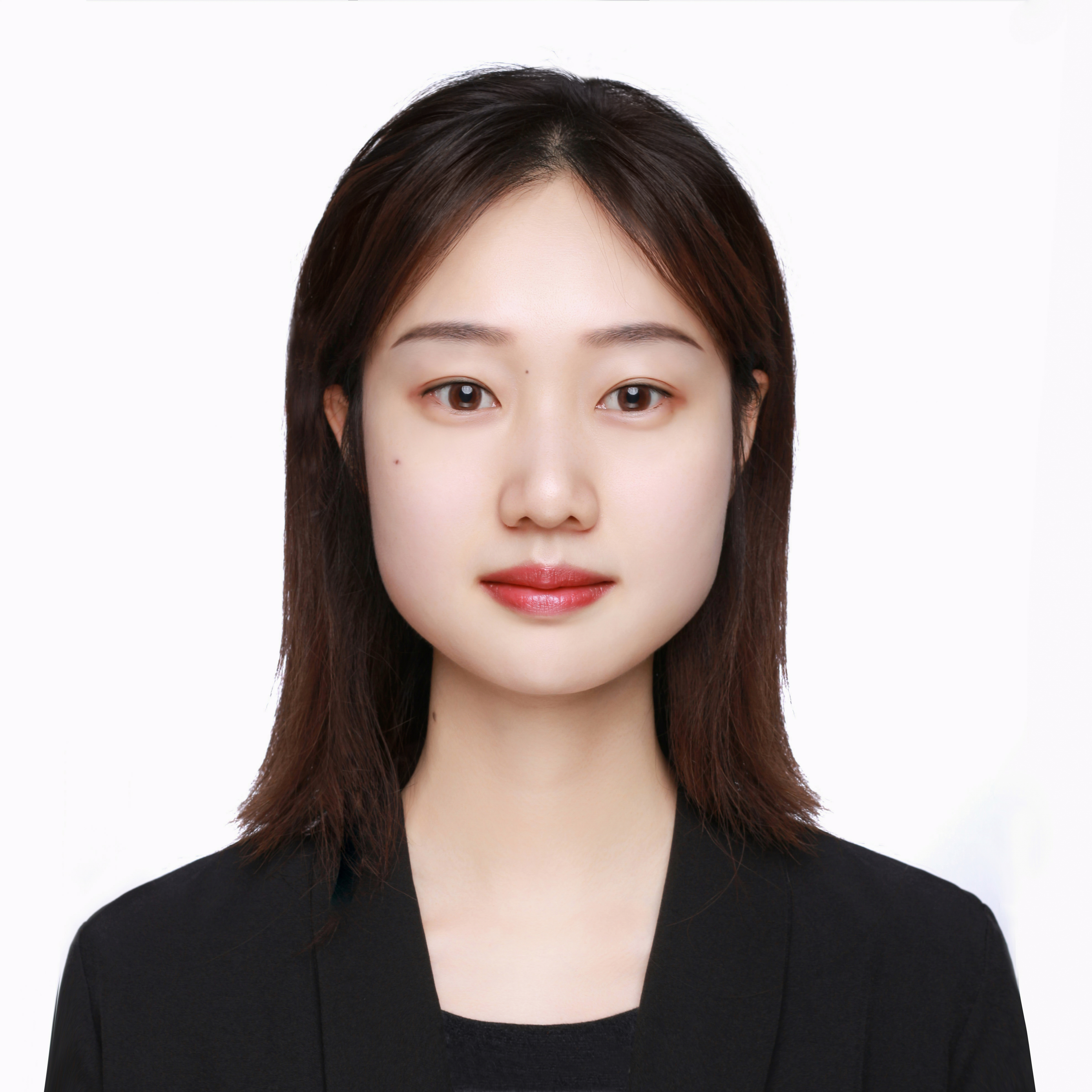}}]{Huake Wang}
received the B.S. degree in electronics engineering, and the Ph.D. degree in signal and information processing from Xidian University, Xi'an, China, in 2015 and 2020, respectively. She was a Visiting Ph.D. Student with the Department of Electrical Engineering, Columbia University, New York, from 2020 to 2021. Currently, she is an Associate Professor with the School of Electronics Engineering, Xidian University. Her research interests include signal processing, new concept radar and intelligent sensing. 
\end{IEEEbiography}

\vspace{-60pt}
\begin{IEEEbiography}[{\includegraphics[width=1in,height=1.25in,clip,keepaspectratio]{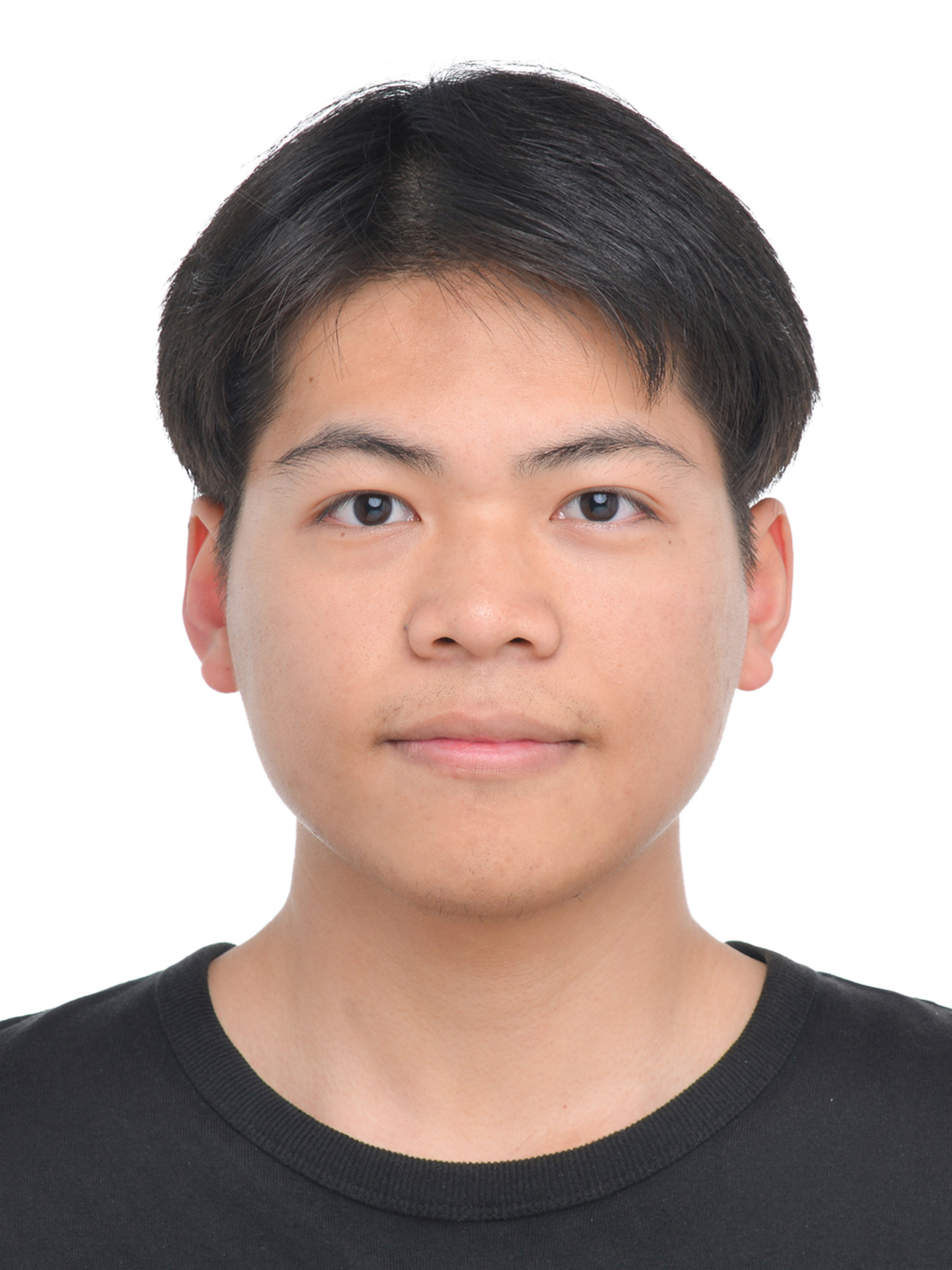}}]{Dongchang Zhang}
received the B.S. degree in Electronic Information Engineering from China JiLiang University in 2023. He is currently pursuing a master's degree in Information and Communication Engineering at Hangzhou Research Institute of Xidian University. His research interests include radar signal processing, jamming suppression and electronic countermeasures. 
\end{IEEEbiography}

\vspace{-60pt}
\begin{IEEEbiography}[{\includegraphics[width=1in,height=1.25in,clip,keepaspectratio]{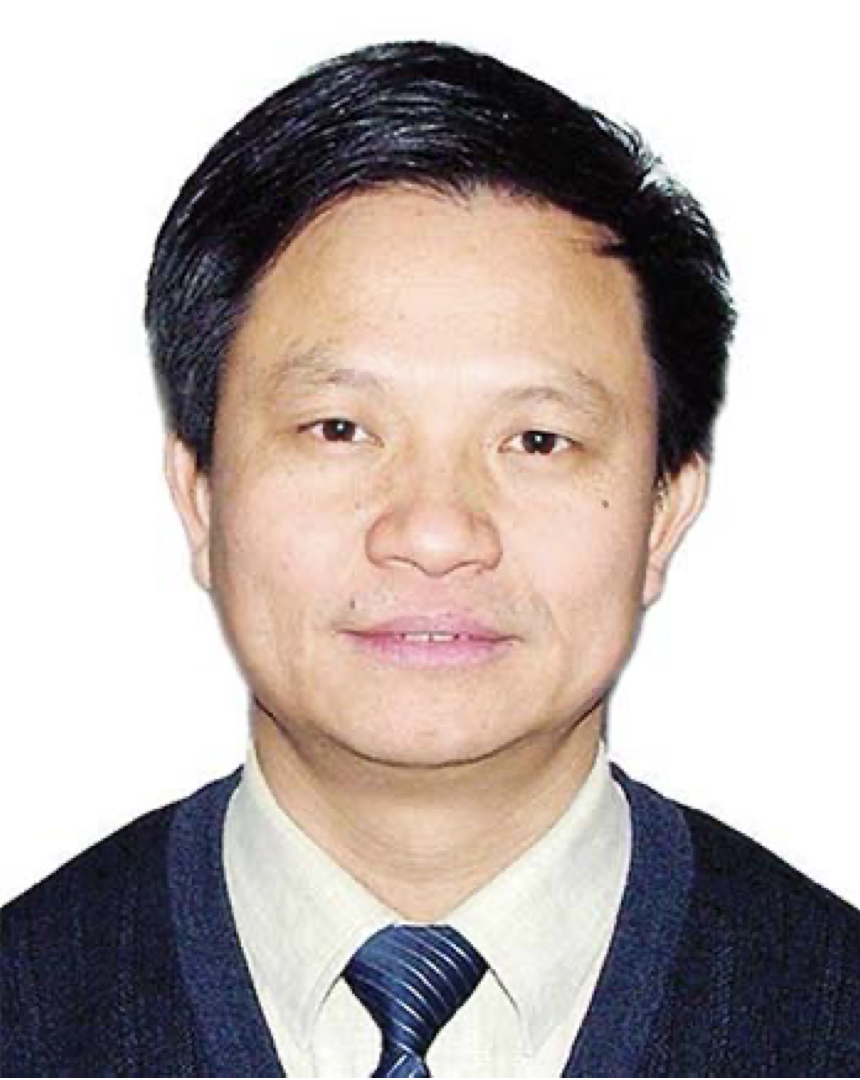}}]{Guisheng Liao}
(Senior Member, IEEE) was born in Guilin, Guangxi, China in 1963. He received the B.S. degree from Guangxi University in mathematics, Guangxi, China, in 1985, and the M.S. degree in computer software and the Ph.D. degree in signal and information processing from Xidian University, Xi’an, China, in 1990, and 1992, respectively. 

He has been a Senior Visiting Scholar with the Chinese University of Hong Kong from 1999 to 2000. He won the National Science Fund for Distinguished Young Scholars in 2008. His research interests include array signal processing, space-time adaptive processing, radar waveform design, and airborne/space surveillance and warning radar systems.

% He is currently a Full Professor with the National Key Laboratory of Radar Signal Processing and served as the 1st Dean with the Hangzhou Institute of Technology, Xidian University since 2021. He has 
% been the Dean with the School of Electronic Engineering, Xidian University from 2013 to 2021. He has been a Senior Visiting Scholar with the Chinese University of Hong Kong from 1999 to 2000. He won the National Science Fund for Distinguished Young Scholars in 2008. His research interests include array signal processing, space-time adaptive processing, radar waveform design, and airborne/space surveillance and warning radar systems.
\end{IEEEbiography}

\vspace{-60pt}
\begin{IEEEbiography}[{\includegraphics[width=1in,height=1.25in,clip,keepaspectratio]{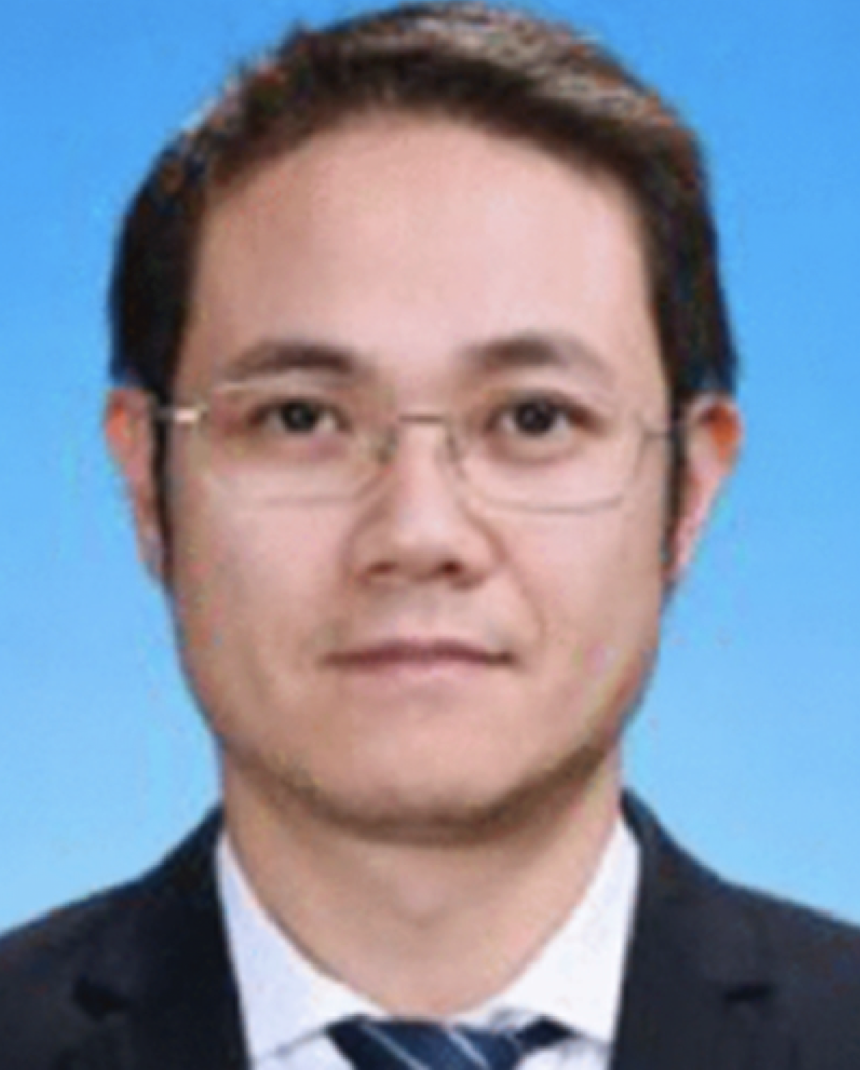}}]{Yinghui Quan}
(Senior Member, IEEE) received the B.S. and Ph.D. degrees in electrical engineering from Xidian University, Xi’an, China, in 2004 and 2012, respectively. 

He is currently a Full Professor with the Department of Remote Sensing Science and Technology, School of Electronic Engineering, Xidian University. His research interests include intelligent sensing and agile radar.
\end{IEEEbiography}

\end{document}